\documentclass[pdflatex, sn-mathphys]{sn-jnl}

\usepackage{graphicx}%
\usepackage{multirow}%
\usepackage{amsmath,amssymb,amsfonts}%
\usepackage[title]{appendix}
\usepackage{xcolor}
\usepackage{textcomp}

\usepackage{ulem}
\usepackage{lmodern}
\usepackage{aas_macros}
\usepackage{color}
\usepackage{hyperref}
\usepackage{multibib} 

\newcites{Methods}{Methods References}

\newcommand{\Ms}{ \mathrm{M}_\odot }

\begin{document}

\title[Apocalypse When? No Certainty of a Milky Way - Andromeda Collision]{Apocalypse When? No Certainty of a Milky Way - Andromeda Collision}

\author*[1,2]{\fnm{Till} \sur{Sawala}}\email{till.sawala@helsinki.fi}
\author[1,3]{\fnm{Jehanne} \sur{Delhomelle}}
\author[2]{\fnm{Alis J.} \sur{Deason}}
\author[2]{\fnm{Carlos S.} \sur{Frenk}}
\author[1]{\fnm{Peter H.} \sur{Johansson}}
\author[1]{\fnm{Atte} \sur{Keitaanranta}}
\author[1]{\fnm{Alexander} \sur{Rawlings}}
\author[1]{\fnm{Ruby} \sur{Wright}}

\affil[1]{\orgdiv{Department of Physics}, \orgname{University of Helsinki}, \orgaddress{\street{Gustaf H\"allstr\"omin katu~2}, \city{Helsinki}, \postcode{FI-00014}, \country{Finland}}}

\affil[2]{\orgdiv{Institute for Computational Cosmology}, \orgname{Durham University}, \orgaddress{\street{South Road}, \city{Durham}, \postcode{DH1 3LE}, \country{United Kingdom}}}

\affil[3]{\orgname{Université Toulouse III - Paul Sabatier}, \orgaddress{\street{118 Route de Narbonne}, \city{Toulouse}, \postcode{F-31062}, \country{France}}}

\abstract{It is commonly believed that our own Milky Way is on a collision course with the neighbouring Andromeda galaxy. As a result of their merger, predicted in around five~billion years, the two large spiral galaxies that define the present Local Group would form a new elliptical galaxy. Here we consider the latest and most accurate observations by the \textit{Gaia} and \textit{Hubble} space telescopes, along with recent consensus mass estimates to derive possible future scenarios and identify the major sources of uncertainty in the evolution of the Local Group over the next 10 billion years. We find that the next most massive Local Group member galaxies -- namely, M33 and the Large Magellanic Cloud -- distinctly and radically affect the Milky Way - Andromeda orbit. While including M33 increases the merger probability, the orbit of the Large Magellanic Cloud runs perpendicular to the Milky Way - Andromeda orbit and makes their merger less likely. In the full system, we find that uncertainties in the present positions, motions, and masses of all galaxies leave room for drastically different outcomes, and a probability of close to 50\% that there is no Milky Way - Andromeda merger during the next 10 billion years.}

\maketitle
\noindent
The Local Group (LG) contains two large spiral galaxies, our own Milky Way (MW) and the Andromeda galaxy (hereafter M31), along with approximately 100 known smaller galaxies \cite{McConnachie-2012}. In addition, it likely hosts additional galaxies yet to be discovered~\cite{Newton-2023} and, according to the standard cosmological model, a vast number of completely dark substructures \cite{Sawala-2016a}.

The negative radial velocity of M31 towards the MW has been known for over a century \cite{Slipher-1913}, even before its distance was first accurately measured  \cite{Hubble-1929}. However, while indirect methods had since been used to constrain the transverse components of M31's velocity vector \cite{Peebles-2001, Loeb-2005, VanderMarel-2008}, direct measurements of the minute proper motions were only achieved much more recently, with the Hubble Space Telescope (HST) \cite{Sohn-2012}.

The first numerical studies \cite{Dubinski-1996} of a possible MW-M31 merger predate even the early estimates of the transverse velocity. The finding that the MW-M31 motion is close to radial immediately led to the prediction of a likely future collision and merger \cite{Cox-Loeb-2008, VanderMarel-2012, VanderMarel-2019}. This scenario has since become the prevalent narrative \cite{Cowen-2012, Harvey-Smith-2020} and textbook knowledge \cite{Binney-Tremaine-2008, Eicher-2015}. 

\subsection*{Predicting the future of the Local Group}
The MW and M31 both contain remnants of past mergers and interactions with other galaxies \cite{Helmi-2018, Belokurov-2018, Ruiz-Lara-2020, Naidu-2021}. Predicting future mergers requires knowledge about the present coordinates, velocities, and masses of the systems partaking in the interaction. In addition to the gravitational force between galaxies, dynamical friction is the dominant process in the lead-up to galactic mergers. It describes a transfer of orbital kinetic energy to internal energy of the objects involved, and consequently leads to the decay of galactic orbits.

In this study, we parameterise the density profile of each halo as isotropic Navarro-Frenk-White profiles \cite{Navarro-1997}, use standard assumptions about their concentrations and velocity dispersion profiles \cite{Zentner-2003} and calculate dynamical friction using an analytic formalism (see Methods for a detailed description). While non-gravitational processes, such as gas drag and star formation leading to increased central densities, etc., also shape the final stages of a merger, the phase of the orbit that defines the occurrence of mergers is largely determined by gravity, which in turn, is dominated by the dark matter component in the standard cosmological model \cite{Cautun-2019}. 

The simplest model for the MW-M31 orbit, as considered by \cite{Cox-Loeb-2008}, contains only the two main galaxies. Due to the planar symmetry, only five parameters are required: the two masses, along with the initial separation and the two velocity components. In general, an orbit with $N>2$ galaxies requires specifying $3 \times (N-1)$ coordinates and $3 \times (N-1)$ velocity components, along with the $N$ masses. More recently,  \cite{VanderMarel-2012} and \cite{VanderMarel-2019} have considered three-body orbits including M33, the third most massive LG galaxy; and \cite{VanderMarel-2019} also considered the Large Magellanic Cloud (LMC), the fourth most massive LG galaxy. They still conclude that a merger is certain. We will consider two-body, three-body, and four-body systems to study the future evolution of the LG and reveal the distinct effects of the M33 and LMC on the MW-M31 orbit.

\subsection*{A fiducial model based on the most accurate values available}
In the context of the Local Group, it is important to note that apart from the sky positions, all parameters including distance moduli, line-of-sight velocities, proper motions, masses, and concentrations carry non-negligible uncertainties. We will use Monte Carlo (NC) sampling of all values to investigate how these observational errors propagate to uncertainties on the future evolution, and in particular, the probability of a merger between the MW and M31.

Our fiducial LG model contains the four most massive LG members: the MW, M31, M33, and the LMC, based on the latest and most accurate available data. The mass of the MW has recently been extensively studied using \textit{Gaia} data, with a consensus emerging of a total mass close to $10^{12} \Ms$. We adopt a value of $M_{200} = 1 \pm 0.2 \times 10^{12} \Ms$ (excluding the mass of the LMC, which we treat separately). For all other galaxies, there is significantly more uncertainty. For M31, we adopt $M_{200} = 1.3 \pm 0.4 \times 10^{12} \Ms$. For M33, we assume $M_{200} = 3 \pm 1 \times 10^{11} \Ms$ and for the LMC we assume $M_{200} = 1.5 \pm 0.5 \times 10^{11} \Ms$. See Methods for a review of mass estimates.

The line-of-sight velocities are well-known and we adopt the values given by \cite{McConnachie-2012}. For the distance moduli, $\mu$, we choose the most accurate and precise recent values in the literature: \cite{Pietrzynski-2019} for the LMC and \cite{Ou-2023} for M33. For M31, we use recent HST Cepheid results \cite{Li-2021}.

For the proper motions, where we use the notation,  $\mu_\delta$ for the proper motion in declination and $\mu_\alpha^* = \mu_\alpha \mbox{cos} \delta$ for the proper motion in right ascension, we use the Hubble Space Telescope (HST) proper motions of \cite{Kallivayalil-2013} for the LMC, and the combined HST and {\it Gaia} DR2 proper motions of \cite{VanderMarel-2019} for M33. Our fiducial M31 proper motions are the most precise values in the literature based on {\it Gaia} DR3 astrometry \cite{Salomon-2021}. However, we find very similar results using the combined {\it Gaia} DR2 and HST proper motions of \cite{VanderMarel-2019}, as shown in Extended Data Figure~\ref{fig:ext-distances-vDM}. All assumed values are listed in Extended Table~\ref{tab:data}. We assume Galactocentric coordinates of RA$_\mathrm{GC}= 266.41^{\circ}$, Dec$_\mathrm{GC}=-28.94^{\circ}$ \cite{Reid-2004}, $d_\mathrm{GC}=8.122$~kpc \cite{Gravity-2018}, and a velocity of $(12.9, 245.6, 7.78)$~kms$^{-1}$ with respect to the Sun \cite{Reid-2004, Gravity-2018, Drimmel-2018}.

To account for the fact that the true probability distributions may not be Gaussian and to exclude possible effects caused by unrealistic or even unphysical outliers, we truncate all distributions at $\pm 2\sigma$. This truncation increases the probability of a MW-M31 merger by only $\sim 10\%$ (see Extended Data Figure~\ref{fig:ext-truncation}). For the fiducial model, we use $50000$ MC samples while for all other variants, we use $2500$ samples, ensuring that all statistical errors on the merger rate are below $1\%$.

\subsection*{The evolution of the MW-M31-M33-LMC system}
\begin{figure*}
\centering
\vspace{-.4cm}
    \includegraphics[height=5.8cm, trim={0cm 0.8cm 0cm 0.6cm},clip]{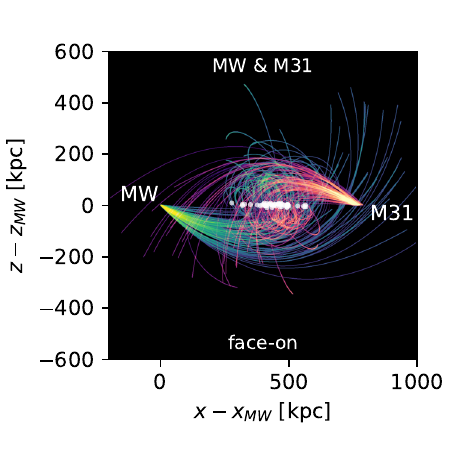}
    \includegraphics[height=5.8cm, trim={1.6cm 0.8cm 0cm 0.6cm},clip]{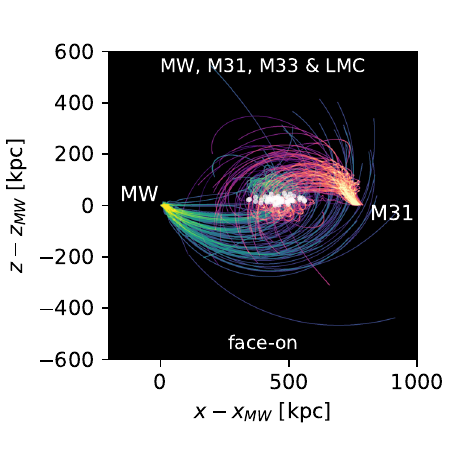} \\
    \includegraphics[height=6.4cm, trim={0cm 0.2cm 0cm 0.6cm},clip]{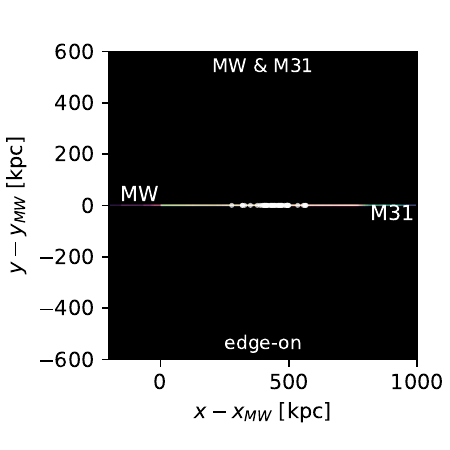} 
    \includegraphics[height=6.4cm, trim={1.6cm 0.2cm 0cm 0.6cm},clip]{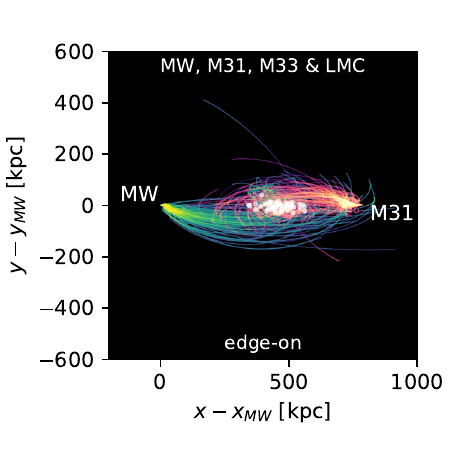}  
    \vspace{.5cm}
    \caption{Possible future MW-M31 orbits. Coloured lines show probability densities for the positions of the MW and M31 in 100 Monte Carlo samples of the fiducial model, integrated over 10 Gyr or until the merger. On the top and bottom panels, respectively, trajectories are projected in the orbital plane and perpendicular to the orbital plane defined by the initial positions and velocities of the MW and M31. White markers denote MW-M31 mergers. In the left column, we show the MW-M31 two-body system, while in the right, we show the four-body system that includes the MW, M31, M33, and the LMC.} 
    \label{fig:orbits}
\end{figure*}

The Monte Carlo initial conditions are integrated numerically using a symplectic direct scheme (see Methods for details). Figure~\ref{fig:orbits} shows 100 realisations each of the MW-M31 orbit, in the two-body MW-M31, and the four-body MW-M31-M33-LMC systems. Figure~\ref{fig:distances} shows the evolution of the distances between the MW and M31 for the same sets of orbits, and additionally, for the MW-M31-M33 and MW-M31-LMC three-body systems.

The MW-M31 two-body orbit evolves in a plane and leads to a merger in slightly less than half of cases, the majority of which occur during the second pericentre. The addition of M33 increases the merger probability to $\sim 2/3$, with a similar median merger time. However, the addition of the LMC has the opposite effect: the pure MW-M31-LMC system experiences a merger in only slightly more than $1/3$ of cases and the merger probability of the full M31-MW-M33-LMC system is just over $50\%$.

For each system, we also show the single orbit obtained by adopting the most likely value of each observable, either in the fiducial model or assuming HST+\textit{Gaia} DR2 proper motions for M31 \cite{VanderMarel-2019}. In the case of the MW-M31-M33 system, which is the one considered by \cite{VanderMarel-2012} and \cite{VanderMarel-2019},
we reproduce similar orbits and times of first pericentre and merger (the differences are likely due to differences in the other model parameters). However, despite using newer and more precise measurements, when we perform a Monte Carlo analysis, we find considerable uncertainty in the outcome that was not previously reported. In particular, in the full MW-M31-M33-LMC system, a merger between the MW and M31 occurs within the next 10 Gyr in only approximately half of the cases. This is in stark contrast to all previous results that had only considered the most likely values without accounting for the numerous and significant uncertainties.

\begin{figure*}
\centering
    \includegraphics[width=6cm, trim={0cm 0.0cm 0cm 0.0cm},clip]{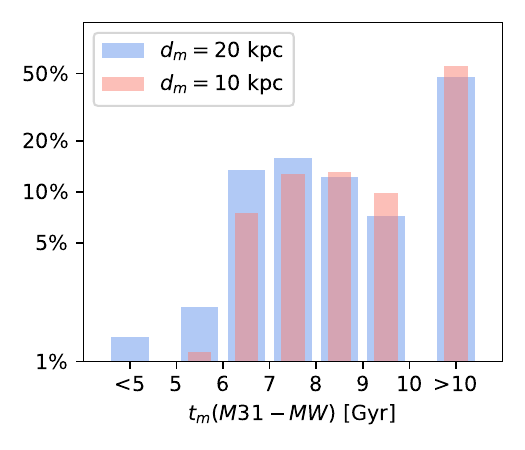}  
    \includegraphics[width=6cm, trim={0cm 0.0cm 0cm 0.0cm},clip]{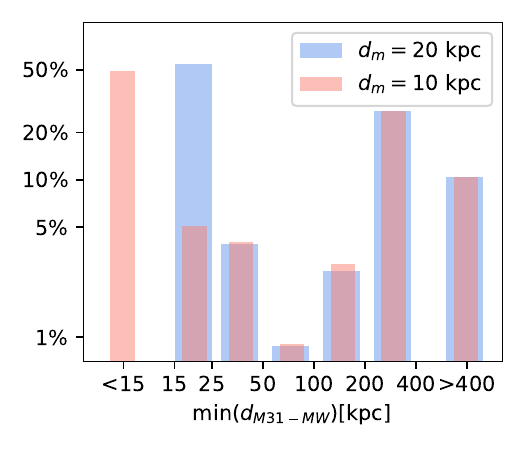} 
    \includegraphics[width=6cm, trim={0cm 0.0cm 0cm 0.0cm},clip]{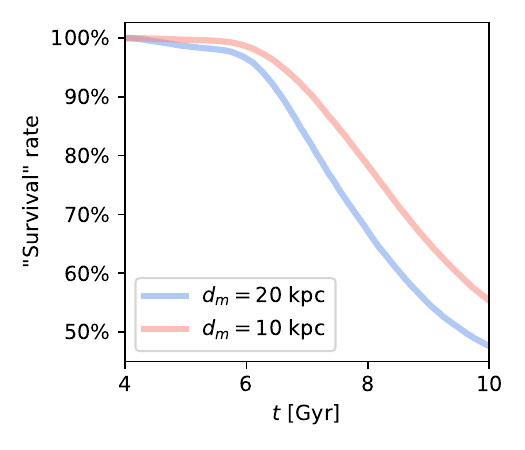} 
    \caption{Distributions of the merger time, $t_m$, the minimum distance, min$(d_{M31-MW})$, and the ``survival'' rate of the MW in the MW-M31-M33-LMC system. Blue and red colours show results using a distance threshold of $20$~kpc (our default) or $10$~kpc for a merger, respectively. Mergers happen on average later when the distance threshold is lower, but the fraction of systems that merge within 10 Gyr is similar. The median time to merger is 7.6 Gyr for systems that merge with a 20~kpc threshold and 8.0 Gyr for systems that merge with a 10~kpc threshold. The distributions of minimum distance show a clear bimodality, independently of the threshold: about half of the systems reach the merger threshold within 10 Gyr while the vast majority of the remaining systems do not approach closer than 200 kpc. The ``survival'' rate of the MW drops sharply between $\sim 6 - 9$~Gyr and levels off afterwards.}
    \label{fig:distributions-1D}
\end{figure*}

In Figure~\ref{fig:distributions-1D}, we show the probability distributions of the merger time and of the minimum distance between the MW and M31, as well as the ``survival" rate of the MW over time, i.e. the probability that no merger with M31 has occurred. In the fiducial model, we consider a merger to occur when the distance between any two galaxies is below 20 kpc, but we find that our results are not sensitive to this particular choice (see Methods for details). Due to the effect of dynamical friction, we find that there are two distinct possibilities for the eventual fate of the MW and M31: orbits that come within less than $\sim 200$ kpc eventually merge, which would likely lead to the formation of an intermediate-mass elliptical galaxy \cite{Dubinski-1996, Forbes-2000, Schiavi-2020}. For systems that merge, we find a median time of $7.6$~Gyr in the fiducial model, or $8.0$~Gyr adopting a 10~kpc threshold. By contrast, orbits with larger pericentres do not decay due to dynamical friction. In this case, the MW and M31 continue to evolve in isolation. Based on the best current data, both outcomes are almost equally likely.

\subsection*{The roles of M33 and the LMC}
\begin{figure*}
\centering
  
  \includegraphics[height=5cm, trim={0.cm 1.6cm 0.6cm 0.5cm},clip]{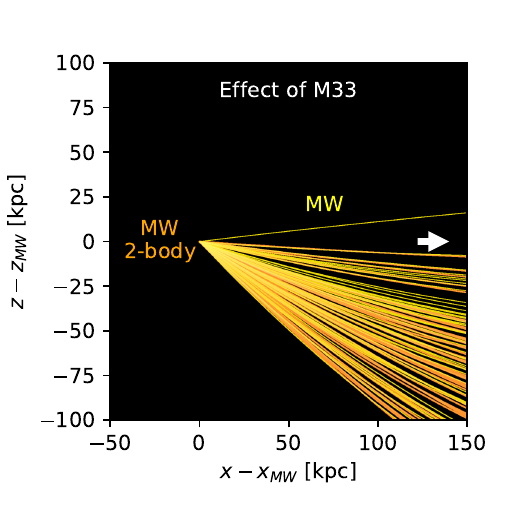} 
  \includegraphics[height=5cm, trim={1cm 1.6cm 0.6cm 0.5cm},clip]{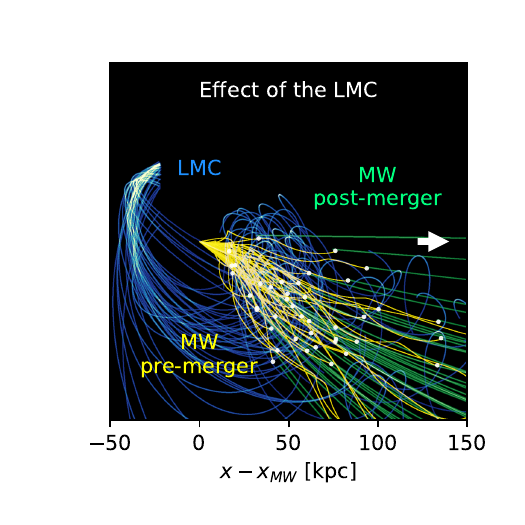} \\
  \includegraphics[height=6.2cm, trim={0cm 0cm 0.6cm 0.5cm},clip]{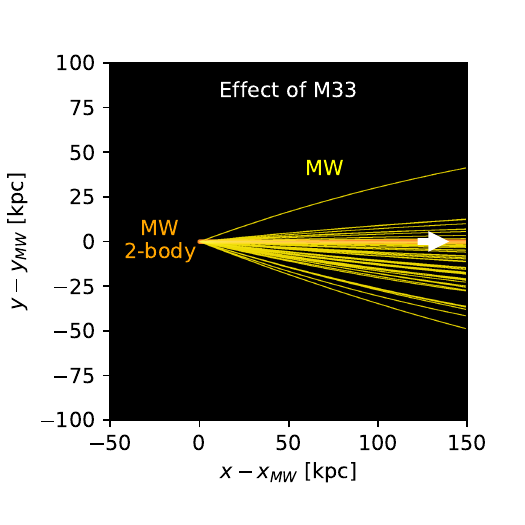} 
  \includegraphics[height=6.2cm, trim={1cm 0cm 0.6cm 0.5cm},clip]{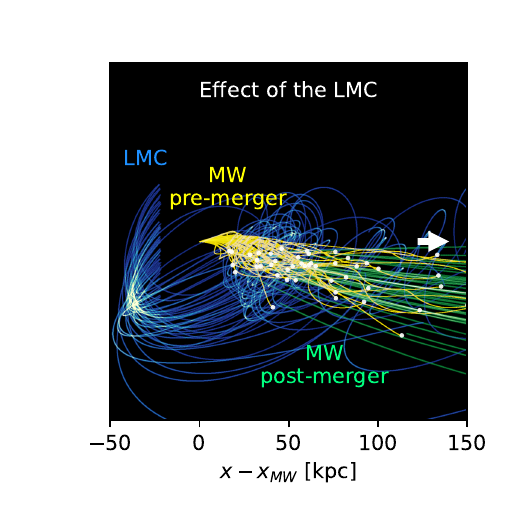} 
  
\caption{Effects of M33 and the LMC on the trajectory of the MW. As in Figures~\ref{fig:orbits} and~\ref{fig:orbits-satellites-M31}, panels in the top row are projected in the orbital plane defined by the initial MW and M31 positions and velocities, while those in the bottom row are projected perpendicular to this orbital plane. Arrows point towards the initial position of M31. On the left, we compare MW trajectories in the two-body MW-M31 system to those in the three-body MW-M31-M33 system. On the right, we show the orbit of the LMC, and that of the MW before and after the merger with the LMC in the MW-M31-LMC system. Compared to the two-body orbit, the inclusion of M33 reduces the transverse velocity of the MW relative to M31 and introduces only a small velocity perpendicular to the MW-M31 orbital plane. By contrast, the LMC increases the MW-M31 transverse velocity and causes a larger velocity perpendicular to the MW-M31 orbital plane.}
\label{fig:orbits-satellites-MW}
\end{figure*}

\begin{figure*}
\centering

  \includegraphics[height=5cm, trim={0.cm 1.6cm 0.6cm 0.5cm},clip]{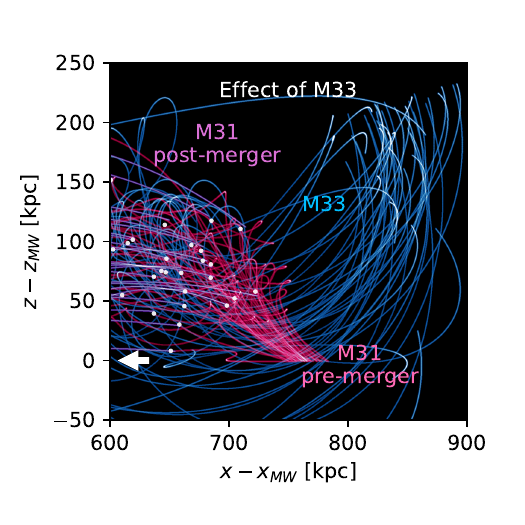} 
  \includegraphics[height=5cm, trim={1cm 1.6cm 0.6cm 0.5cm},clip]{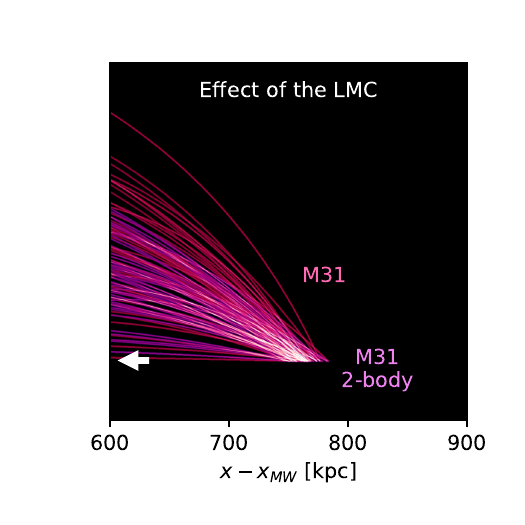} \\
  \includegraphics[height=6.2cm, trim={0.cm 0cm 0.6cm 0.5cm},clip]{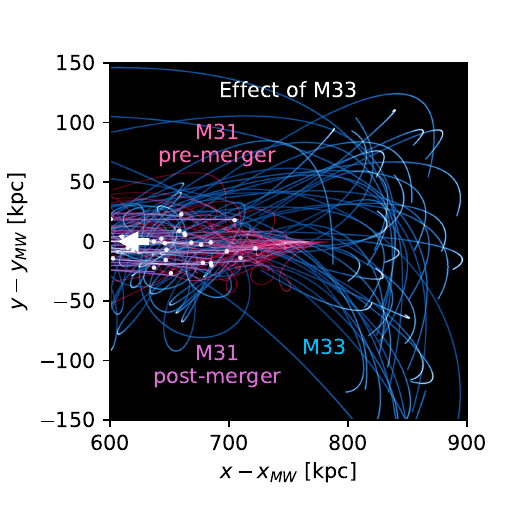} 
  \includegraphics[height=6.2cm, trim={1cm 0cm 0.6cm 0.5cm},clip]{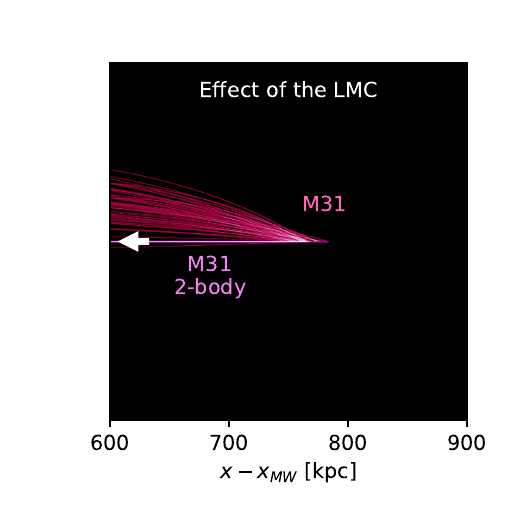}

\caption{Effects of M33 and the LMC on the trajectory of M31. As in Figures~\ref{fig:orbits} and~\ref{fig:orbits-satellites-MW}, panels in the top row are projected in the orbital plane defined by the initial MW and M31 positions and velocities, while those in the bottom row are projected perpendicular to this orbital plane. Arrows point towards the initial position of the MW. On the right, we compare M31 trajectories in the two-body MW-M31 system to those in the three-body MW-M31-LMC system. On the left, we show the orbit of M33, and that of M31 before and after a possible merger with M33 in the MW-M31-M33 system. Compared to the 2-body orbit, the inclusion of the LMC increases the transverse velocity of M31 and introduces motion perpendicular to the initial orbital plane. On the other hand, the inclusion of M33 reduces the transverse velocity of M33 with respect to the MW.}
\label{fig:orbits-satellites-M31}
\end{figure*}

The distinct effects of each satellite on the MW-M31 orbit are illustrated in Figures~\ref{fig:orbits-satellites-MW} and~\ref{fig:orbits-satellites-M31}, where we compare the trajectories of the MW and M31 in two-body systems and in systems that also include either M33 or the LMC. Both satellites provide some additional acceleration in the radial direction of the MW-M31 orbit. However, importantly, both satellites also affect the motion of their respective hosts. Including M33 in the calculation decreases the transverse velocity of M31 with respect to the MW. By contrast, as already pointed out in \cite{Penarrubia-2016}, at its current orbital phase, the recoil due to the LMC results in a lower transverse velocity measured between the MW and M31. During its short orbital period of $\sim 1.5$ Gyr, the LMC will accelerate the MW to a higher transverse velocity. In addition, the inclusion of M33 largely provides momentum in the original MW-M31 plane, while the inclusion of the LMC also provides significant momentum perpendicular to the MW-M31 plane. In our analysis, the LMC is certain to merge with the MW, and M33 is highly likely $(\sim 86\%)$ to merge with M31 before any possible MW-M31 merger. The net effect of adding M33 to the two-body system is to increase the merger probability, while the net effect of adding the LMC is to decrease it.

\subsection*{Sources of uncertainty}
\begin{figure*}
    \includegraphics[width=1.65in, trim={0cm 0.35cm 0cm 0.35cm},clip]{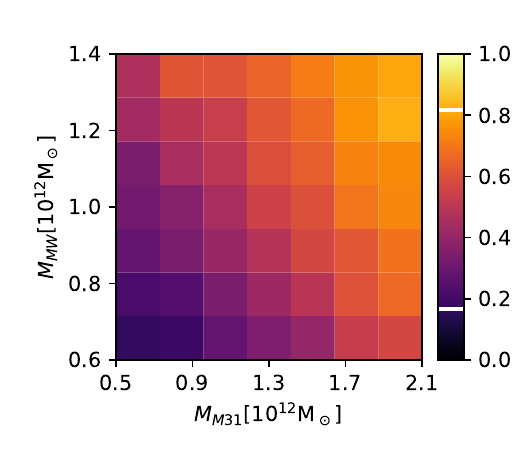}  
    \includegraphics[width=1.65in, trim={0cm 0.35cm 0cm 0.35cm},clip]{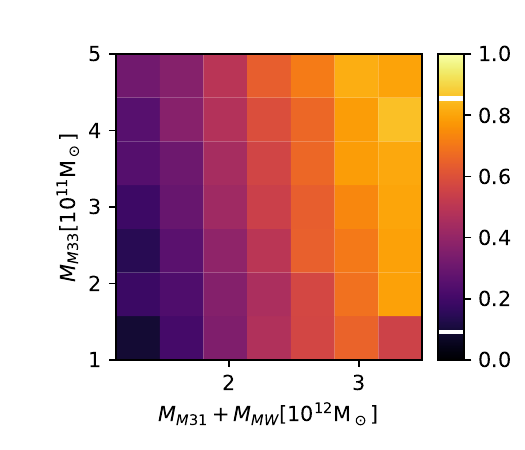}  
    \includegraphics[width=1.65in, trim={0cm 0.35cm 0cm 0.35cm},clip]{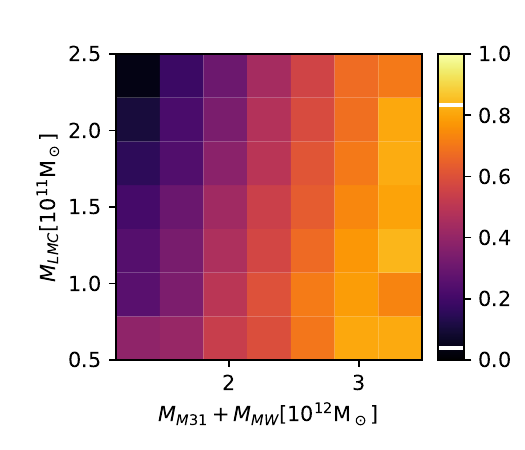}    \\
    \includegraphics[width=1.65in, trim={0cm 0.35cm 0cm 0.35cm},clip]{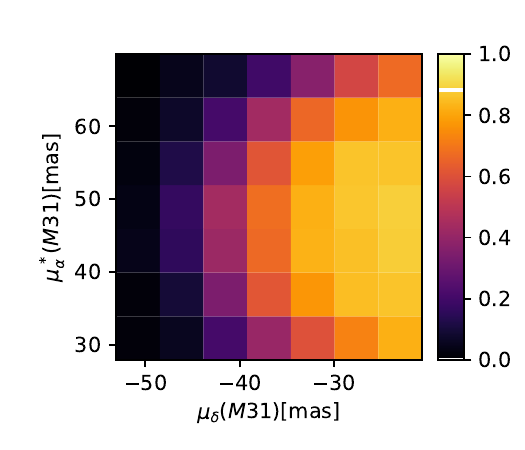} 
    \includegraphics[width=1.65in, trim={0cm 0.35cm 0cm 0.35cm},clip]{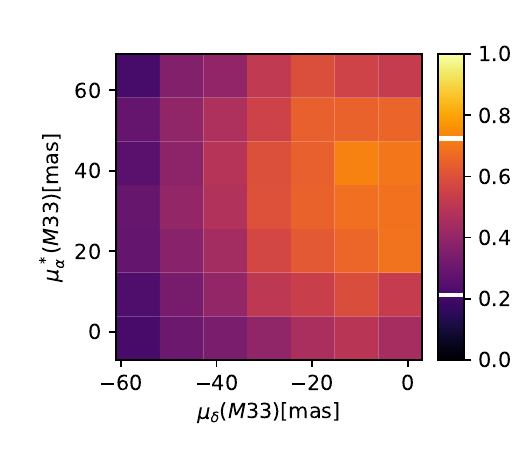} 
    \includegraphics[width=1.65in, trim={0cm 0.35cm 0cm 0.35cm},clip]{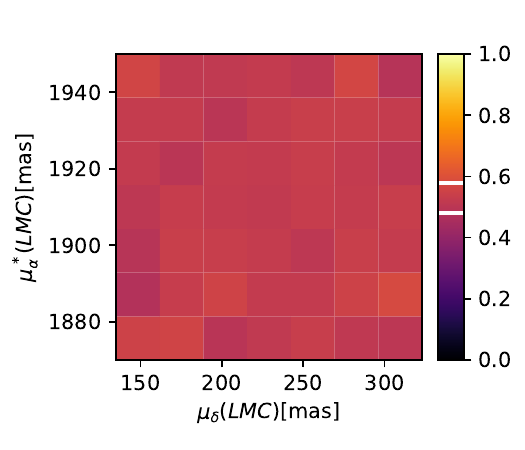}    \\
    \includegraphics[width=1.65in, trim={0cm 0.35cm 0cm 0.35cm},clip]{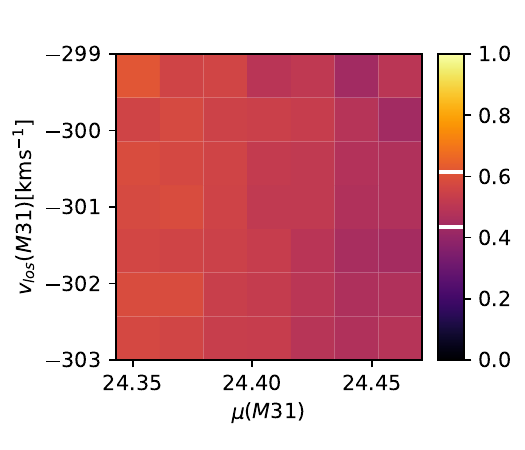} 
    \includegraphics[width=1.65in, trim={0cm 0.35cm 0cm 0.35cm},clip]{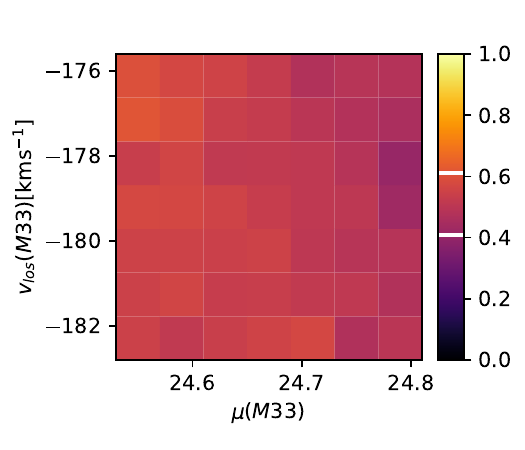} 
    \includegraphics[width=1.65in, trim={0cm 0.35cm 0cm 0.35cm},clip]{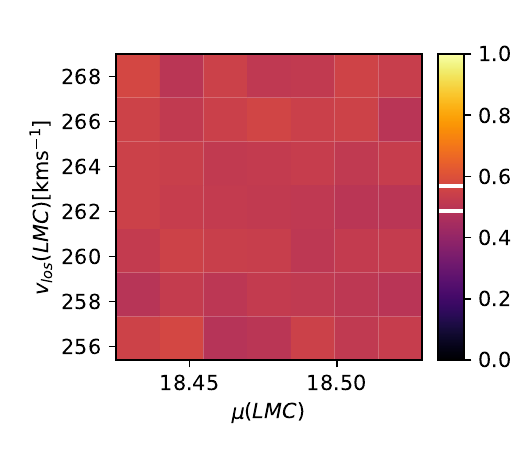} \\
    \caption{Dependence of the merger probability on observables of the MW-M31-M33-LMC system. In each panel, we show the probability of a MW-M31 merger within 10 Gyr as a function of two variables, with all other observables sampling the probability distributions of the fiducial model. White lines on the colour bars indicate the minimum and maximum merger probabilities for the range of values shown on each panel, indicating the sensitivity of the merger probability on the two corresponding variables. In the top row, we show the dependence on different masses, the second row shows the dependence on proper motions, and the third row shows the dependence on distance moduli and line of sight velocities. The axes extend to $\pm 2 \sigma$ of the fiducial model. The merger probability is positively correlated with the mass of the MW, M31, and M33, and negatively correlated with the mass of the LMC. The merger probability strongly depends on $\mu_\delta (M31)$ and $\mu_\alpha^* (M31)$, but also varies significantly with $\mu_\delta (M33)$ and $\mu_\alpha^* (M33)$. The uncertainties in the distance moduli for M31 and M33 also contribute to the uncertainty of the outcome, while the effect of the line-of-sight velocities is small.}
    \label{fig:distributions}
\end{figure*}

In Figure~\ref{fig:distributions}, we show how the merger probability depends on the different observables. In each panel, we show the dependence on two variables in the $\pm 2 \sigma$ ranges, with the remaining variables Monte Carlo sampled. The effects of the concentration parameters are shown in Extended Data Figure~\ref{fig:ext-concentration}. We find that all masses, the proper motions of M31 and M33, and the distance moduli of M31 and M33, significantly impact the probability of a merger. The merger probability is positively correlated with the masses of the MW, M31, and M33 and negatively correlated with the mass of the LMC. The impact of the satellite masses is more pronounced for lower combined masses of the MW and M31. The $\pm 2 \sigma$ ranges of the M31 proper motions include values that imply a merger probability above $90\%$, but also values that imply a merger probability close to zero. The most likely proper motions (assuming no errors) only lead to a merger in $\sim 2/3$ of cases.

Future more precise proper motion measurements may significantly change the expected outcome, although they could make the merger either more or less likely. However, if they fall within $\pm 1 \sigma$ of the current most likely values, even precise M31 proper motion measurements alone will not suffice to determine the outcome.

Even the comparatively high precision of the line-of-sight velocity and distance moduli for M33 and M31 contribute significant uncertainty, with the probability of a merger varying between $40\%$ and over $60\%$ for different combinations in the $\pm 2 \sigma$ ranges around the most likely values.

Given the considerable measurement errors, it is worth noting that cosmological simulations result in a present-day MW-M31 transverse velocity prior of $v_t = 75_{-40}^{+65}$~kms$^{-1}$ \cite{Sawala-2023a}. There is thus no reason to assume that the transverse velocity measured using Gaia DR3 (most likely value of $v_t=76$~kms$^{-1}$) is overestimated or to expect that more precise measurements will result in a lower value. It is also worth noting that a perfectly radial present-day M31-MW motion ($\mu_\alpha^* = 38.03$~mas~yr$^{-1}$, $\mu_\delta = -21.37$~mas~yr$^{-1}$) is neither compatible with current observations nor does it result in the highest merger probability in the four-body system.

\subsection*{Summary}
Even using the latest and most precise available observational data, the future evolution of the Local Group is uncertain. Intriguingly, we find an almost equal probability for the widely publicised merger scenario (albeit with a later median time to merger) and one where the MW and M31 survive unscathed. We reach this conclusion by including the LMC and importantly, for the first time, considering the relevant uncertainties in the observables.

Our results are not sensitive to the necessary choices of gravitational softening (see Extended Data Figure~\ref{fig:ext-softening}), merger threshold (Extended Data Figure~\ref{fig:ext-threshold}), or dynamical friction scheme (Extended Data Figure~\ref{fig:ext-dynamical-friction})).

While we have shown that considerable uncertainty results from the proper motion measurements of M31, we also find that a more accurate prediction requires more precise measurements of the positions, motions, and masses of all participating galaxies. The dependence of the evolution of the MW-M31 system on the treatment of other LG galaxies points to further uncertainties. Cosmological simulations suggest that $\sim 25\%$ of the bound mass of the LG is outside the two main haloes \cite{Sawala-2023}. The next most massive individual Local Group galaxies that could impact the MW-M31 orbit are M32 (an M31 satellite and possible merger remnant \cite{DSouza-2018}) and the Small Magellanic Cloud (SMC), a satellite of the MW. Both are at least a factor of five less massive than the LMC, and we find that including the SMC, for which proper motion measurements are available, does not significantly change the MW-M31 merger probability (see Extended Data Figure~\ref{fig:ext-SMC}). However, the unaccounted cumulative effects of additional substructures as well as that of the cosmic environment introduce further uncertainty, particularly towards the far future. An accurate prediction of the evolution even from perfectly precise observations may require cosmological constrained simulations \cite{Libeskind-2020, Sawala-2022, Wempe-2024} that can account for these effects. Meanwhile, the assumptions and simplifications we have made here are likely conservative regarding our central claim, that there is considerable uncertainty about the MW-M31 merger.

Upcoming {\it Gaia} data releases will improve the proper motion constraints and mass models are continuously refined. However, it is clear that Galactic eschatology is still in its infancy and significant work is required before the eventual fate of the Local Group can be predicted with any certainty. As it stands, proclamations of the impending demise of our Galaxy appear greatly exaggerated.
\clearpage

\bibliography{main} 

\begin{thebibliography}{10}
\expandafter\ifx\csname url\endcsname\relax
  \def\url#1{\texttt{#1}}\fi
\expandafter\ifx\csname urlprefix\endcsname\relax\def\urlprefix{URL }\fi
\providecommand{\bibinfo}[2]{#2}
\providecommand{\eprint}[2][]{\url{#2}}

\bibitem{White-1976}
\bibinfo{author}{{White}, S. D.~M.}
\newblock \bibinfo{title}{{A note on the minimum impact parameter for dynamical friction involving spherical clusters}}.
\newblock \emph{\bibinfo{journal}{\mnras}} \textbf{\bibinfo{volume}{174}}, \bibinfo{pages}{467--470} (\bibinfo{year}{1976}).

\bibitem{Jethwa-2016}
\bibinfo{author}{{Jethwa}, P.}, \bibinfo{author}{{Erkal}, D.} \& \bibinfo{author}{{Belokurov}, V.}
\newblock \bibinfo{title}{{A Magellanic origin of the DES dwarfs}}.
\newblock \emph{\bibinfo{journal}{\mnras}} \textbf{\bibinfo{volume}{461}}, \bibinfo{pages}{2212--2233} (\bibinfo{year}{2016}).
\newblock \eprint{1603.04420}.

\bibitem{Boylan-Kolchin-2008}
\bibinfo{author}{{Boylan-Kolchin}, M.}, \bibinfo{author}{{Ma}, C.-P.} \& \bibinfo{author}{{Quataert}, E.}
\newblock \bibinfo{title}{{Dynamical friction and galaxy merging time-scales}}.
\newblock \emph{\bibinfo{journal}{\mnras}} \textbf{\bibinfo{volume}{383}}, \bibinfo{pages}{93--101} (\bibinfo{year}{2008}).
\newblock \eprint{0707.2960}.

\bibitem{Karl-2010}
\bibinfo{author}{{Karl}, S.~J.} \emph{et~al.}
\newblock \bibinfo{title}{{One Moment in Time{\textemdash}Modeling Star Formation in the Antennae}}.
\newblock \emph{\bibinfo{journal}{\apjl}} \textbf{\bibinfo{volume}{715}}, \bibinfo{pages}{L88--L93} (\bibinfo{year}{2010}).
\newblock \eprint{1003.0685}.

\bibitem{Lahen-2018}
\bibinfo{author}{{Lah{\'e}n}, N.}, \bibinfo{author}{{Johansson}, P.~H.}, \bibinfo{author}{{Rantala}, A.}, \bibinfo{author}{{Naab}, T.} \& \bibinfo{author}{{Frigo}, M.}
\newblock \bibinfo{title}{{The fate of the Antennae galaxies}}.
\newblock \emph{\bibinfo{journal}{\mnras}} \textbf{\bibinfo{volume}{475}}, \bibinfo{pages}{3934--3958} (\bibinfo{year}{2018}).
\newblock \eprint{1709.00010}.

\bibitem{Callingham-2019}
\bibinfo{author}{{Callingham}, T.~M.} \emph{et~al.}
\newblock \bibinfo{title}{{The mass of the Milky Way from satellite dynamics}}.
\newblock \emph{\bibinfo{journal}{\mnras}} \textbf{\bibinfo{volume}{484}}, \bibinfo{pages}{5453--5467} (\bibinfo{year}{2019}).
\newblock \eprint{1808.10456}.

\bibitem{Fritz-2020}
\bibinfo{author}{{Fritz}, T.~K.}, \bibinfo{author}{{Di Cintio}, A.}, \bibinfo{author}{{Battaglia}, G.}, \bibinfo{author}{{Brook}, C.} \& \bibinfo{author}{{Taibi}, S.}
\newblock \bibinfo{title}{{The mass of our Galaxy from satellite proper motions in the Gaia era}}.
\newblock \emph{\bibinfo{journal}{\mnras}} \textbf{\bibinfo{volume}{494}}, \bibinfo{pages}{5178--5193} (\bibinfo{year}{2020}).
\newblock \eprint{2001.02651}.

\bibitem{Li-2020}
\bibinfo{author}{{Li}, Z.-Z.} \emph{et~al.}
\newblock \bibinfo{title}{{Constraining the Milky Way Mass Profile with Phase-space Distribution of Satellite Galaxies}}.
\newblock \emph{\bibinfo{journal}{\apj}} \textbf{\bibinfo{volume}{894}}, \bibinfo{pages}{10} (\bibinfo{year}{2020}).
\newblock \eprint{1912.02086}.

\bibitem{Rodriguez-Wimberly-2022}
\bibinfo{author}{{Rodriguez Wimberly}, M.~K.} \emph{et~al.}
\newblock \bibinfo{title}{{Sizing from the smallest scales: the mass of the Milky Way}}.
\newblock \emph{\bibinfo{journal}{\mnras}} \textbf{\bibinfo{volume}{513}}, \bibinfo{pages}{4968--4982} (\bibinfo{year}{2022}).
\newblock \eprint{2109.00633}.

\bibitem{Cautun-2020}
\bibinfo{author}{{Cautun}, M.} \emph{et~al.}
\newblock \bibinfo{title}{{The milky way total mass profile as inferred from Gaia DR2}}.
\newblock \emph{\bibinfo{journal}{\mnras}} \textbf{\bibinfo{volume}{494}}, \bibinfo{pages}{4291--4313} (\bibinfo{year}{2020}).
\newblock \eprint{1911.04557}.

\bibitem{Karukes-2020}
\bibinfo{author}{{Karukes}, E.~V.}, \bibinfo{author}{{Benito}, M.}, \bibinfo{author}{{Iocco}, F.}, \bibinfo{author}{{Trotta}, R.} \& \bibinfo{author}{{Geringer-Sameth}, A.}
\newblock \bibinfo{title}{{A robust estimate of the Milky Way mass from rotation curve data}}.
\newblock \emph{\bibinfo{journal}{"Journal of Cosmology and Astroparticle Physics"}} \textbf{\bibinfo{volume}{2020}}, \bibinfo{pages}{033} (\bibinfo{year}{2020}).
\newblock \eprint{1912.04296}.

\bibitem{Ablimit-2020}
\bibinfo{author}{{Ablimit}, I.}, \bibinfo{author}{{Zhao}, G.}, \bibinfo{author}{{Flynn}, C.} \& \bibinfo{author}{{Bird}, S.~A.}
\newblock \bibinfo{title}{{The Rotation Curve, Mass Distribution, and Dark Matter Content of the Milky Way from Classical Cepheids}}.
\newblock \emph{\bibinfo{journal}{\apjl}} \textbf{\bibinfo{volume}{895}}, \bibinfo{pages}{L12} (\bibinfo{year}{2020}).
\newblock \eprint{2004.13768}.

\bibitem{Shen-2022}
\bibinfo{author}{{Shen}, J.} \emph{et~al.}
\newblock \bibinfo{title}{{The Mass of the Milky Way from the H3 Survey}}.
\newblock \emph{\bibinfo{journal}{\apj}} \textbf{\bibinfo{volume}{925}}, \bibinfo{pages}{1} (\bibinfo{year}{2022}).
\newblock \eprint{2111.09327}.

\bibitem{Watkins-2019}
\bibinfo{author}{{Watkins}, L.~L.}, \bibinfo{author}{{van der Marel}, R.~P.}, \bibinfo{author}{{Sohn}, S.~T.} \& \bibinfo{author}{{Evans}, N.~W.}
\newblock \bibinfo{title}{{Evidence for an Intermediate-mass Milky Way from Gaia DR2 Halo Globular Cluster Motions}}.
\newblock \emph{\bibinfo{journal}{\apj}} \textbf{\bibinfo{volume}{873}}, \bibinfo{pages}{118} (\bibinfo{year}{2019}).
\newblock \eprint{1804.11348}.

\bibitem{Prudil-2022}
\bibinfo{author}{{Prudil}, Z.} \emph{et~al.}
\newblock \bibinfo{title}{{Milky Way archaeology using RR Lyrae and type II Cepheids II. High velocity RR Lyrae stars, and mass of the Milky Way}}.
\newblock \emph{\bibinfo{journal}{arXiv e-prints}} \bibinfo{pages}{arXiv:2206.00417} (\bibinfo{year}{2022}).
\newblock \eprint{2206.00417}.

\bibitem{Slizewski-2022}
\bibinfo{author}{{Slizewski}, A.} \emph{et~al.}
\newblock \bibinfo{title}{{Galactic Mass Estimates Using Dwarf Galaxies as Kinematic Tracers}}.
\newblock \emph{\bibinfo{journal}{\apj}} \textbf{\bibinfo{volume}{924}}, \bibinfo{pages}{131} (\bibinfo{year}{2022}).
\newblock \eprint{2108.12474}.

\bibitem{Wang-2020}
\bibinfo{author}{{Wang}, W.}, \bibinfo{author}{{Han}, J.}, \bibinfo{author}{{Cautun}, M.}, \bibinfo{author}{{Li}, Z.} \& \bibinfo{author}{{Ishigaki}, M.~N.}
\newblock \bibinfo{title}{{The mass of our Milky Way}}.
\newblock \emph{\bibinfo{journal}{Science China Physics, Mechanics, and Astronomy}} \textbf{\bibinfo{volume}{63}}, \bibinfo{pages}{109801} (\bibinfo{year}{2020}).
\newblock \eprint{1912.02599}.

\bibitem{Zhang-2024}
\bibinfo{author}{{Zhang}, X.}, \bibinfo{author}{{Chen}, B.}, \bibinfo{author}{{Chen}, P.}, \bibinfo{author}{{Sun}, J.} \& \bibinfo{author}{{Tian}, Z.}
\newblock \bibinfo{title}{{The rotation curve and mass distribution of M31}}.
\newblock \emph{\bibinfo{journal}{\mnras}} \textbf{\bibinfo{volume}{528}}, \bibinfo{pages}{2653--2666} (\bibinfo{year}{2024}).
\newblock \eprint{2401.01517}.

\bibitem{Watkins-2010}
\bibinfo{author}{{Watkins}, L.~L.}, \bibinfo{author}{{Evans}, N.~W.} \& \bibinfo{author}{{An}, J.~H.}
\newblock \bibinfo{title}{{The masses of the Milky Way and Andromeda galaxies}}.
\newblock \emph{\bibinfo{journal}{\mnras}} \textbf{\bibinfo{volume}{406}}, \bibinfo{pages}{264--278} (\bibinfo{year}{2010}).
\newblock \eprint{1002.4565}.

\bibitem{Tollerud-2012}
\bibinfo{author}{{Tollerud}, E.~J.} \emph{et~al.}
\newblock \bibinfo{title}{{The SPLASH Survey: Spectroscopy of 15 M31 Dwarf Spheroidal Satellite Galaxies}}.
\newblock \emph{\bibinfo{journal}{\apj}} \textbf{\bibinfo{volume}{752}}, \bibinfo{pages}{45} (\bibinfo{year}{2012}).
\newblock \eprint{1112.1067}.

\bibitem{Tamm-2012}
\bibinfo{author}{{Tamm}, A.}, \bibinfo{author}{{Tempel}, E.}, \bibinfo{author}{{Tenjes}, P.}, \bibinfo{author}{{Tihhonova}, O.} \& \bibinfo{author}{{Tuvikene}, T.}
\newblock \bibinfo{title}{{Stellar mass map and dark matter distribution in M 31}}.
\newblock \emph{\bibinfo{journal}{\aap}} \textbf{\bibinfo{volume}{546}}, \bibinfo{pages}{A4} (\bibinfo{year}{2012}).
\newblock \eprint{1208.5712}.

\bibitem{Chemin-2009}
\bibinfo{author}{{Chemin}, L.}, \bibinfo{author}{{Carignan}, C.} \& \bibinfo{author}{{Foster}, T.}
\newblock \bibinfo{title}{{H I Kinematics and Dynamics of Messier 31}}.
\newblock \emph{\bibinfo{journal}{\apj}} \textbf{\bibinfo{volume}{705}}, \bibinfo{pages}{1395--1415} (\bibinfo{year}{2009}).
\newblock \eprint{0909.3846}.

\bibitem{Fardal-2013}
\bibinfo{author}{{Fardal}, M.~A.} \emph{et~al.}
\newblock \bibinfo{title}{{Inferring the Andromeda Galaxy's mass from its giant southern stream with Bayesian simulation sampling}}.
\newblock \emph{\bibinfo{journal}{\mnras}} \textbf{\bibinfo{volume}{434}}, \bibinfo{pages}{2779--2802} (\bibinfo{year}{2013}).
\newblock \eprint{1307.3219}.

\bibitem{Veljanoski-2013}
\bibinfo{author}{{Veljanoski}, J.} \emph{et~al.}
\newblock \bibinfo{title}{{Kinematics of Outer Halo Globular Clusters in M31}}.
\newblock \emph{\bibinfo{journal}{\apjl}} \textbf{\bibinfo{volume}{768}}, \bibinfo{pages}{L33} (\bibinfo{year}{2013}).
\newblock \eprint{1303.7368}.

\bibitem{Veljanoski-2014}
\bibinfo{author}{{Veljanoski}, J.} \emph{et~al.}
\newblock \bibinfo{title}{{The outer halo globular cluster system of M31 - II. Kinematics}}.
\newblock \emph{\bibinfo{journal}{\mnras}} \textbf{\bibinfo{volume}{442}}, \bibinfo{pages}{2929--2950} (\bibinfo{year}{2014}).
\newblock \eprint{1406.0186}.

\bibitem{Kafle-2018}
\bibinfo{author}{{Kafle}, P.~R.}, \bibinfo{author}{{Sharma}, S.}, \bibinfo{author}{{Lewis}, G.~F.}, \bibinfo{author}{{Robotham}, A. S.~G.} \& \bibinfo{author}{{Driver}, S.~P.}
\newblock \bibinfo{title}{{The need for speed: escape velocity and dynamical mass measurements of the Andromeda galaxy}}.
\newblock \emph{\bibinfo{journal}{\mnras}} \textbf{\bibinfo{volume}{475}}, \bibinfo{pages}{4043--4054} (\bibinfo{year}{2018}).
\newblock \eprint{1801.03949}.

\bibitem{Sofue-2015}
\bibinfo{author}{{Sofue}, Y.}
\newblock \bibinfo{title}{{Dark halos of M 31 and the Milky Way}}.
\newblock \emph{\bibinfo{journal}{\pasj}} \textbf{\bibinfo{volume}{67}}, \bibinfo{pages}{75} (\bibinfo{year}{2015}).
\newblock \eprint{1504.05368}.

\bibitem{Corbelli-2000}
\bibinfo{author}{{Corbelli}, E.} \& \bibinfo{author}{{Salucci}, P.}
\newblock \bibinfo{title}{{The extended rotation curve and the dark matter halo of M33}}.
\newblock \emph{\bibinfo{journal}{\mnras}} \textbf{\bibinfo{volume}{311}}, \bibinfo{pages}{441--447} (\bibinfo{year}{2000}).
\newblock \eprint{astro-ph/9909252}.

\bibitem{Corbelli-2003}
\bibinfo{author}{{Corbelli}, E.}
\newblock \bibinfo{title}{{Dark matter and visible baryons in M33}}.
\newblock \emph{\bibinfo{journal}{\mnras}} \textbf{\bibinfo{volume}{342}}, \bibinfo{pages}{199--207} (\bibinfo{year}{2003}).
\newblock \eprint{astro-ph/0302318}.

\bibitem{Kam-2017}
\bibinfo{author}{{Kam}, S.~Z.} \emph{et~al.}
\newblock \bibinfo{title}{{H I Kinematics and Mass Distribution of Messier 33}}.
\newblock \emph{\bibinfo{journal}{\aj}} \textbf{\bibinfo{volume}{154}}, \bibinfo{pages}{41} (\bibinfo{year}{2017}).

\bibitem{Guo-2010}
\bibinfo{author}{{Guo}, Q.}, \bibinfo{author}{{White}, S.}, \bibinfo{author}{{Li}, C.} \& \bibinfo{author}{{Boylan-Kolchin}, M.}
\newblock \bibinfo{title}{{How do galaxies populate dark matter haloes?}}
\newblock \emph{\bibinfo{journal}{\mnras}} \textbf{\bibinfo{volume}{404}}, \bibinfo{pages}{1111--1120} (\bibinfo{year}{2010}).
\newblock \eprint{0909.4305}.

\bibitem{Patel-2017}
\bibinfo{author}{{Patel}, E.}, \bibinfo{author}{{Besla}, G.} \& \bibinfo{author}{{Sohn}, S.~T.}
\newblock \bibinfo{title}{{Orbits of massive satellite galaxies - I. A close look at the Large Magellanic Cloud and a new orbital history for M33}}.
\newblock \emph{\bibinfo{journal}{\mnras}} \textbf{\bibinfo{volume}{464}}, \bibinfo{pages}{3825--3849} (\bibinfo{year}{2017}).
\newblock \eprint{1609.04823}.

\bibitem{Patel-2018}
\bibinfo{author}{{Patel}, E.}, \bibinfo{author}{{Carlin}, J.~L.}, \bibinfo{author}{{Tollerud}, E.~J.}, \bibinfo{author}{{Collins}, M. L.~M.} \& \bibinfo{author}{{Dooley}, G.~A.}
\newblock \bibinfo{title}{{{\ensuremath{\Lambda}}CDM predictions for the satellite population of M33}}.
\newblock \emph{\bibinfo{journal}{\mnras}} \textbf{\bibinfo{volume}{480}}, \bibinfo{pages}{1883--1897} (\bibinfo{year}{2018}).
\newblock \eprint{1807.05318}.

\bibitem{Erkal-2019}
\bibinfo{author}{{Erkal}, D.} \emph{et~al.}
\newblock \bibinfo{title}{{The total mass of the Large Magellanic Cloud from its perturbation on the Orphan stream}}.
\newblock \emph{\bibinfo{journal}{\mnras}} \textbf{\bibinfo{volume}{487}}, \bibinfo{pages}{2685--2700} (\bibinfo{year}{2019}).
\newblock \eprint{1812.08192}.

\bibitem{Vasiliev-2021}
\bibinfo{author}{{Vasiliev}, E.}, \bibinfo{author}{{Belokurov}, V.} \& \bibinfo{author}{{Erkal}, D.}
\newblock \bibinfo{title}{{Tango for three: Sagittarius, LMC, and the Milky Way}}.
\newblock \emph{\bibinfo{journal}{\mnras}} \textbf{\bibinfo{volume}{501}}, \bibinfo{pages}{2279--2304} (\bibinfo{year}{2021}).
\newblock \eprint{2009.10726}.

\bibitem{Shipp-2021}
\bibinfo{author}{{Shipp}, N.} \emph{et~al.}
\newblock \bibinfo{title}{{Measuring the Mass of the Large Magellanic Cloud with Stellar Streams Observed by S $^{5}$}}.
\newblock \emph{\bibinfo{journal}{\apj}} \textbf{\bibinfo{volume}{923}}, \bibinfo{pages}{149} (\bibinfo{year}{2021}).
\newblock \eprint{2107.13004}.

\bibitem{Erkal-2021}
\bibinfo{author}{{Erkal}, D.} \emph{et~al.}
\newblock \bibinfo{title}{{Detection of the LMC-induced sloshing of the Galactic halo}}.
\newblock \emph{\bibinfo{journal}{\mnras}} \textbf{\bibinfo{volume}{506}}, \bibinfo{pages}{2677--2684} (\bibinfo{year}{2021}).
\newblock \eprint{2010.13789}.

\bibitem{Erkal-2020}
\bibinfo{author}{{Erkal}, D.} \& \bibinfo{author}{{Belokurov}, V.~A.}
\newblock \bibinfo{title}{{Limit on the LMC mass from a census of its satellites}}.
\newblock \emph{\bibinfo{journal}{\mnras}} \textbf{\bibinfo{volume}{495}}, \bibinfo{pages}{2554--2563} (\bibinfo{year}{2020}).
\newblock \eprint{1907.09484}.

\bibitem{Correa-Magnus-2022}
\bibinfo{author}{{Correa Magnus}, L.} \& \bibinfo{author}{{Vasiliev}, E.}
\newblock \bibinfo{title}{{Measuring the Milky Way mass distribution in the presence of the LMC}}.
\newblock \emph{\bibinfo{journal}{\mnras}} \textbf{\bibinfo{volume}{511}}, \bibinfo{pages}{2610--2630} (\bibinfo{year}{2022}).
\newblock \eprint{2110.00018}.

\bibitem{Watkins-2024}
\bibinfo{author}{{Watkins}, L.~L.}, \bibinfo{author}{{van der Marel}, R.~P.} \& \bibinfo{author}{{Bennet}, P.}
\newblock \bibinfo{title}{{The Mass of the Large Magellanic Cloud from the Three-dimensional Kinematics of Its Globular Clusters}}.
\newblock \emph{\bibinfo{journal}{\apj}} \textbf{\bibinfo{volume}{963}}, \bibinfo{pages}{84} (\bibinfo{year}{2024}).
\newblock \eprint{2401.14458}.

\bibitem{Vasiliev-2023}
\bibinfo{author}{{Vasiliev}, E.}
\newblock \bibinfo{title}{{The Effect of the LMC on the Milky Way System}}.
\newblock \emph{\bibinfo{journal}{Galaxies}} \textbf{\bibinfo{volume}{11}}, \bibinfo{pages}{59} (\bibinfo{year}{2023}).
\newblock \eprint{2304.09136}.

\bibitem{Neto-2007}
\bibinfo{author}{{Neto}, A.~F.} \emph{et~al.}
\newblock \bibinfo{title}{{The statistics of {\ensuremath{\Lambda}} CDM halo concentrations}}.
\newblock \emph{\bibinfo{journal}{\mnras}} \textbf{\bibinfo{volume}{381}}, \bibinfo{pages}{1450--1462} (\bibinfo{year}{2007}).
\newblock \eprint{0706.2919}.

\bibitem{Ludlow-2014}
\bibinfo{author}{{Ludlow}, A.~D.} \emph{et~al.}
\newblock \bibinfo{title}{{The mass-concentration-redshift relation of cold dark matter haloes}}.
\newblock \emph{\bibinfo{journal}{\mnras}} \textbf{\bibinfo{volume}{441}}, \bibinfo{pages}{378--388} (\bibinfo{year}{2014}).
\newblock \eprint{1312.0945}.

\bibitem{Wang-2015}
\bibinfo{author}{{Wang}, W.} \emph{et~al.}
\newblock \bibinfo{title}{{Estimating the dark matter halo mass of our Milky Way using dynamical tracers}}.
\newblock \emph{\bibinfo{journal}{\mnras}} \textbf{\bibinfo{volume}{453}}, \bibinfo{pages}{377--400} (\bibinfo{year}{2015}).
\newblock \eprint{1502.03477}.

\bibitem{Correa-2015}
\bibinfo{author}{{Correa}, C.~A.}, \bibinfo{author}{{Wyithe}, J. S.~B.}, \bibinfo{author}{{Schaye}, J.} \& \bibinfo{author}{{Duffy}, A.~R.}
\newblock \bibinfo{title}{{The accretion history of dark matter haloes - III. A physical model for the concentration-mass relation}}.
\newblock \emph{\bibinfo{journal}{\mnras}} \textbf{\bibinfo{volume}{452}}, \bibinfo{pages}{1217--1232} (\bibinfo{year}{2015}).
\newblock \eprint{1502.00391}.

\bibitem{Diemer-2015}
\bibinfo{author}{{Diemer}, B.} \& \bibinfo{author}{{Kravtsov}, A.~V.}
\newblock \bibinfo{title}{{A Universal Model for Halo Concentrations}}.
\newblock \emph{\bibinfo{journal}{\apj}} \textbf{\bibinfo{volume}{799}}, \bibinfo{pages}{108} (\bibinfo{year}{2015}).
\newblock \eprint{1407.4730}.

\bibitem{Watkins-2013}
\bibinfo{author}{{Watkins}, L.~L.}, \bibinfo{author}{{Evans}, N.~W.} \& \bibinfo{author}{{van de Ven}, G.}
\newblock \bibinfo{title}{{A census of orbital properties of the M31 satellites}}.
\newblock \emph{\bibinfo{journal}{\mnras}} \textbf{\bibinfo{volume}{430}}, \bibinfo{pages}{971--985} (\bibinfo{year}{2013}).
\newblock \eprint{1211.2638}.

\bibitem{Graczyk-2014}
\bibinfo{author}{{Graczyk}, D.} \emph{et~al.}
\newblock \bibinfo{title}{{The Araucaria Project. The Distance to the Small Magellanic Cloud from Late-type Eclipsing Binaries}}.
\newblock \emph{\bibinfo{journal}{\apj}} \textbf{\bibinfo{volume}{780}}, \bibinfo{pages}{59} (\bibinfo{year}{2014}).
\newblock \eprint{1311.2340}.

\bibitem{Harris-2006}
\bibinfo{author}{{Harris}, J.} \& \bibinfo{author}{{Zaritsky}, D.}
\newblock \bibinfo{title}{{Spectroscopic Survey of Red Giants in the Small Magellanic Cloud. I. Kinematics}}.
\newblock \emph{\bibinfo{journal}{\aj}} \textbf{\bibinfo{volume}{131}}, \bibinfo{pages}{2514--2524} (\bibinfo{year}{2006}).
\newblock \eprint{astro-ph/0601025}.

\end{thebibliography}


\begin{thebibliography}{10}
\expandafter\ifx\csname url\endcsname\relax
  \def\url#1{\texttt{#1}}\fi
\expandafter\ifx\csname urlprefix\endcsname\relax\def\urlprefix{URL }\fi
\providecommand{\bibinfo}[2]{#2}
\providecommand{\eprint}[2][]{\url{#2}}

\bibitem{McConnachie-2012}
\bibinfo{author}{{McConnachie}, A.~W.}
\newblock \bibinfo{title}{{The Observed Properties of Dwarf Galaxies in and around the Local Group}}.
\newblock \emph{\bibinfo{journal}{\aj}} \textbf{\bibinfo{volume}{144}}, \bibinfo{pages}{4} (\bibinfo{year}{2012}).
\newblock \eprint{1204.1562}.

\bibitem{Newton-2023}
\bibinfo{author}{{Newton}, O.} \emph{et~al.}
\newblock \bibinfo{title}{{The Undiscovered Ultradiffuse Galaxies of the Local Group}}.
\newblock \emph{\bibinfo{journal}{\apjl}} \textbf{\bibinfo{volume}{946}}, \bibinfo{pages}{L37} (\bibinfo{year}{2023}).
\newblock \eprint{2212.05066}.

\bibitem{Sawala-2016a}
\bibinfo{author}{{Sawala}, T.} \emph{et~al.}
\newblock \bibinfo{title}{{The APOSTLE simulations: solutions to the Local Group's cosmic puzzles}}.
\newblock \emph{\bibinfo{journal}{\mnras}} \textbf{\bibinfo{volume}{457}}, \bibinfo{pages}{1931--1943} (\bibinfo{year}{2016}).
\newblock \eprint{1511.01098}.

\bibitem{Slipher-1913}
\bibinfo{author}{{Slipher}, V.~M.}
\newblock \bibinfo{title}{{The radial velocity of the Andromeda Nebula}}.
\newblock \emph{\bibinfo{journal}{Lowell Observatory Bulletin}} \textbf{\bibinfo{volume}{2}}, \bibinfo{pages}{56--57} (\bibinfo{year}{1913}).

\bibitem{Hubble-1929}
\bibinfo{author}{{Hubble}, E.~P.}
\newblock \bibinfo{title}{{A spiral nebula as a stellar system, Messier 31.}}
\newblock \emph{\bibinfo{journal}{\apj}} \textbf{\bibinfo{volume}{69}}, \bibinfo{pages}{103--158} (\bibinfo{year}{1929}).

\bibitem{Peebles-2001}
\bibinfo{author}{{Peebles}, P.~J.~E.}, \bibinfo{author}{{Phelps}, S.~D.}, \bibinfo{author}{{Shaya}, E.~J.} \& \bibinfo{author}{{Tully}, R.~B.}
\newblock \bibinfo{title}{{Radial and Transverse Velocities of Nearby Galaxies}}.
\newblock \emph{\bibinfo{journal}{\apj}} \textbf{\bibinfo{volume}{554}}, \bibinfo{pages}{104--113} (\bibinfo{year}{2001}).
\newblock \eprint{astro-ph/0010480}.

\bibitem{Loeb-2005}
\bibinfo{author}{{Loeb}, A.}, \bibinfo{author}{{Reid}, M.~J.}, \bibinfo{author}{{Brunthaler}, A.} \& \bibinfo{author}{{Falcke}, H.}
\newblock \bibinfo{title}{{Constraints on the Proper Motion of the Andromeda Galaxy Based on the Survival of Its Satellite M33}}.
\newblock \emph{\bibinfo{journal}{\apj}} \textbf{\bibinfo{volume}{633}}, \bibinfo{pages}{894--898} (\bibinfo{year}{2005}).
\newblock \eprint{astro-ph/0506609}.

\bibitem{VanderMarel-2008}
\bibinfo{author}{{van der Marel}, R.~P.} \& \bibinfo{author}{{Guhathakurta}, P.}
\newblock \bibinfo{title}{{M31 Transverse Velocity and Local Group Mass from Satellite Kinematics}}.
\newblock \emph{\bibinfo{journal}{\apj}} \textbf{\bibinfo{volume}{678}}, \bibinfo{pages}{187--199} (\bibinfo{year}{2008}).
\newblock \eprint{0709.3747}.

\bibitem{Sohn-2012}
\bibinfo{author}{{Sohn}, S.~T.}, \bibinfo{author}{{Anderson}, J.} \& \bibinfo{author}{{van der Marel}, R.~P.}
\newblock \bibinfo{title}{{The M31 Velocity Vector. I. Hubble Space Telescope Proper-motion Measurements}}.
\newblock \emph{\bibinfo{journal}{\apj}} \textbf{\bibinfo{volume}{753}}, \bibinfo{pages}{7} (\bibinfo{year}{2012}).
\newblock \eprint{1205.6863}.

\bibitem{Dubinski-1996}
\bibinfo{author}{{Dubinski}, J.}, \bibinfo{author}{{Mihos}, J.~C.} \& \bibinfo{author}{{Hernquist}, L.}
\newblock \bibinfo{title}{{Using Tidal Tails to Probe Dark Matter Halos}}.
\newblock \emph{\bibinfo{journal}{\apj}} \textbf{\bibinfo{volume}{462}}, \bibinfo{pages}{576} (\bibinfo{year}{1996}).
\newblock \eprint{astro-ph/9509010}.

\bibitem{Cox-Loeb-2008}
\bibinfo{author}{{Cox}, T.~J.} \& \bibinfo{author}{{Loeb}, A.}
\newblock \bibinfo{title}{{The collision between the Milky Way and Andromeda}}.
\newblock \emph{\bibinfo{journal}{\mnras}} \textbf{\bibinfo{volume}{386}}, \bibinfo{pages}{461--474} (\bibinfo{year}{2008}).
\newblock \eprint{0705.1170}.

\bibitem{VanderMarel-2012}
\bibinfo{author}{{van der Marel}, R.~P.} \emph{et~al.}
\newblock \bibinfo{title}{{The M31 Velocity Vector. II. Radial Orbit toward the Milky Way and Implied Local Group Mass}}.
\newblock \emph{\bibinfo{journal}{\apj}} \textbf{\bibinfo{volume}{753}}, \bibinfo{pages}{8} (\bibinfo{year}{2012}).
\newblock \eprint{1205.6864}.

\bibitem{VanderMarel-2019}
\bibinfo{author}{{van der Marel}, R.~P.} \emph{et~al.}
\newblock \bibinfo{title}{{First Gaia Dynamics of the Andromeda System: DR2 Proper Motions, Orbits, and Rotation of M31 and M33}}.
\newblock \emph{\bibinfo{journal}{\apj}} \textbf{\bibinfo{volume}{872}}, \bibinfo{pages}{24} (\bibinfo{year}{2019}).
\newblock \eprint{1805.04079}.

\bibitem{Cowen-2012}
\bibinfo{author}{{Cowen}, R.}
\newblock \bibinfo{title}{{Andromeda on collision course with the Milky Way}}.
\newblock \emph{\bibinfo{journal}{\nat}}  (\bibinfo{year}{2012}).

\bibitem{Harvey-Smith-2020}
\bibinfo{author}{{Harvey-Smith}, L.}
\newblock \emph{\bibinfo{title}{{When galaxies collide}}} (\bibinfo{publisher}{{Melbourne University Press}}, \bibinfo{year}{2020}).

\bibitem{Binney-Tremaine-2008}
\bibinfo{author}{{Binney}, J.} \& \bibinfo{author}{{Tremaine}, S.}
\newblock \emph{\bibinfo{title}{{Galactic Dynamics: Second Edition}}} (\bibinfo{publisher}{{Princeton University Press}}, \bibinfo{year}{2008}).

\bibitem{Eicher-2015}
\bibinfo{author}{Eicher, D.~J.}
\newblock \emph{\bibinfo{title}{The New Cosmos: Answering Astronomy’s Big Questions}} (\bibinfo{publisher}{Cambridge University Press}, \bibinfo{year}{2015}).

\bibitem{Helmi-2018}
\bibinfo{author}{{Helmi}, A.} \emph{et~al.}
\newblock \bibinfo{title}{{The merger that led to the formation of the Milky Way's inner stellar halo and thick disk}}.
\newblock \emph{\bibinfo{journal}{\nat}} \textbf{\bibinfo{volume}{563}}, \bibinfo{pages}{85--88} (\bibinfo{year}{2018}).
\newblock \eprint{1806.06038}.

\bibitem{Belokurov-2018}
\bibinfo{author}{{Belokurov}, V.}, \bibinfo{author}{{Erkal}, D.}, \bibinfo{author}{{Evans}, N.~W.}, \bibinfo{author}{{Koposov}, S.~E.} \& \bibinfo{author}{{Deason}, A.~J.}
\newblock \bibinfo{title}{{Co-formation of the disc and the stellar halo}}.
\newblock \emph{\bibinfo{journal}{\mnras}} \textbf{\bibinfo{volume}{478}}, \bibinfo{pages}{611--619} (\bibinfo{year}{2018}).
\newblock \eprint{1802.03414}.

\bibitem{Ruiz-Lara-2020}
\bibinfo{author}{{Ruiz-Lara}, T.}, \bibinfo{author}{{Gallart}, C.}, \bibinfo{author}{{Bernard}, E.~J.} \& \bibinfo{author}{{Cassisi}, S.}
\newblock \bibinfo{title}{{The recurrent impact of the Sagittarius dwarf on the star formation history of the Milky Way}}.
\newblock \emph{\bibinfo{journal}{Nature Astronomy}} \textbf{\bibinfo{volume}{4}}, \bibinfo{pages}{965--973} (\bibinfo{year}{2020}).
\newblock \eprint{2003.12577}.

\bibitem{Naidu-2021}
\bibinfo{author}{{Naidu}, R.~P.} \emph{et~al.}
\newblock \bibinfo{title}{{Reconstructing the Last Major Merger of the Milky Way with the H3 Survey}}.
\newblock \emph{\bibinfo{journal}{\apj}} \textbf{\bibinfo{volume}{923}}, \bibinfo{pages}{92} (\bibinfo{year}{2021}).
\newblock \eprint{2103.03251}.

\bibitem{Navarro-1997}
\bibinfo{author}{{Navarro}, J.~F.}, \bibinfo{author}{{Frenk}, C.~S.} \& \bibinfo{author}{{White}, S. D.~M.}
\newblock \bibinfo{title}{{A Universal Density Profile from Hierarchical Clustering}}.
\newblock \emph{\bibinfo{journal}{\apj}} \textbf{\bibinfo{volume}{490}}, \bibinfo{pages}{493--508} (\bibinfo{year}{1997}).
\newblock \eprint{astro-ph/9611107}.

\bibitem{Zentner-2003}
\bibinfo{author}{{Zentner}, A.~R.} \& \bibinfo{author}{{Bullock}, J.~S.}
\newblock \bibinfo{title}{{Halo Substructure and the Power Spectrum}}.
\newblock \emph{\bibinfo{journal}{\apj}} \textbf{\bibinfo{volume}{598}}, \bibinfo{pages}{49--72} (\bibinfo{year}{2003}).
\newblock \eprint{astro-ph/0304292}.

\bibitem{Cautun-2019}
\bibinfo{author}{{Cautun}, M.}, \bibinfo{author}{{Deason}, A.~J.}, \bibinfo{author}{{Frenk}, C.~S.} \& \bibinfo{author}{{McAlpine}, S.}
\newblock \bibinfo{title}{{The aftermath of the Great Collision between our Galaxy and the Large Magellanic Cloud}}.
\newblock \emph{\bibinfo{journal}{\mnras}} \textbf{\bibinfo{volume}{483}}, \bibinfo{pages}{2185--2196} (\bibinfo{year}{2019}).
\newblock \eprint{1809.09116}.

\bibitem{Pietrzynski-2019}
\bibinfo{author}{{Pietrzy{\'n}ski}, G.} \emph{et~al.}
\newblock \bibinfo{title}{{A distance to the Large Magellanic Cloud that is precise to one per cent}}.
\newblock \emph{\bibinfo{journal}{\nat}} \textbf{\bibinfo{volume}{567}}, \bibinfo{pages}{200--203} (\bibinfo{year}{2019}).
\newblock \eprint{1903.08096}.

\bibitem{Ou-2023}
\bibinfo{author}{{Ou}, J.-Y.} \emph{et~al.}
\newblock \bibinfo{title}{{A Distance Measurement to M33 Using Optical Photometry of Mira Variables}}.
\newblock \emph{\bibinfo{journal}{\aj}} \textbf{\bibinfo{volume}{165}}, \bibinfo{pages}{137} (\bibinfo{year}{2023}).
\newblock \eprint{2302.02901}.

\bibitem{Li-2021}
\bibinfo{author}{{Li}, S.} \emph{et~al.}
\newblock \bibinfo{title}{{A Sub-2\% Distance to M31 from Photometrically Homogeneous Near-infrared Cepheid Period-Luminosity Relations Measured with the Hubble Space Telescope}}.
\newblock \emph{\bibinfo{journal}{\apj}} \textbf{\bibinfo{volume}{920}}, \bibinfo{pages}{84} (\bibinfo{year}{2021}).
\newblock \eprint{2107.08029}.

\bibitem{Kallivayalil-2013}
\bibinfo{author}{{Kallivayalil}, N.}, \bibinfo{author}{{van der Marel}, R.~P.}, \bibinfo{author}{{Besla}, G.}, \bibinfo{author}{{Anderson}, J.} \& \bibinfo{author}{{Alcock}, C.}
\newblock \bibinfo{title}{{Third-epoch Magellanic Cloud Proper Motions. I. Hubble Space Telescope/WFC3 Data and Orbit Implications}}.
\newblock \emph{\bibinfo{journal}{\apj}} \textbf{\bibinfo{volume}{764}}, \bibinfo{pages}{161} (\bibinfo{year}{2013}).
\newblock \eprint{1301.0832}.

\bibitem{Salomon-2021}
\bibinfo{author}{{Salomon}, J.~B.} \emph{et~al.}
\newblock \bibinfo{title}{{The proper motion of Andromeda from Gaia EDR3: confirming a nearly radial orbit}}.
\newblock \emph{\bibinfo{journal}{\mnras}} \textbf{\bibinfo{volume}{507}}, \bibinfo{pages}{2592--2601} (\bibinfo{year}{2021}).
\newblock \eprint{2012.09204}.

\bibitem{Reid-2004}
\bibinfo{author}{{Reid}, M.~J.} \& \bibinfo{author}{{Brunthaler}, A.}
\newblock \bibinfo{title}{{The Proper Motion of Sagittarius A*. II. The Mass of Sagittarius A*}}.
\newblock \emph{\bibinfo{journal}{\apj}} \textbf{\bibinfo{volume}{616}}, \bibinfo{pages}{872--884} (\bibinfo{year}{2004}).
\newblock \eprint{astro-ph/0408107}.

\bibitem{Gravity-2018}
\bibinfo{author}{{GRAVITY Collaboration}} \emph{et~al.}
\newblock \bibinfo{title}{{Detection of the gravitational redshift in the orbit of the star S2 near the Galactic centre massive black hole}}.
\newblock \emph{\bibinfo{journal}{\aap}} \textbf{\bibinfo{volume}{615}}, \bibinfo{pages}{L15} (\bibinfo{year}{2018}).
\newblock \eprint{1807.09409}.

\bibitem{Drimmel-2018}
\bibinfo{author}{{Drimmel}, R.} \& \bibinfo{author}{{Poggio}, E.}
\newblock \bibinfo{title}{{On the Solar Velocity}}.
\newblock \emph{\bibinfo{journal}{Research Notes of the American Astronomical Society}} \textbf{\bibinfo{volume}{2}}, \bibinfo{pages}{210} (\bibinfo{year}{2018}).

\bibitem{Forbes-2000}
\bibinfo{author}{{Forbes}, D.~A.}, \bibinfo{author}{{Masters}, K.~L.}, \bibinfo{author}{{Minniti}, D.} \& \bibinfo{author}{{Barmby}, P.}
\newblock \bibinfo{title}{{The elliptical galaxy formerly known as the Local Group: merging the globular cluster systems}}.
\newblock \emph{\bibinfo{journal}{\aap}} \textbf{\bibinfo{volume}{358}}, \bibinfo{pages}{471--480} (\bibinfo{year}{2000}).
\newblock \eprint{astro-ph/0001477}.

\bibitem{Schiavi-2020}
\bibinfo{author}{{Schiavi}, R.}, \bibinfo{author}{{Capuzzo-Dolcetta}, R.}, \bibinfo{author}{{Arca Sedda}, M.} \& \bibinfo{author}{{Spera}, M.}
\newblock \bibinfo{title}{{The collision between the Milky Way and Andromeda and the fate of their Supermassive Black Holes}}.
\newblock In \bibinfo{editor}{{Bragaglia}, A.}, \bibinfo{editor}{{Davies}, M.}, \bibinfo{editor}{{Sills}, A.} \& \bibinfo{editor}{{Vesperini}, E.} (eds.) \emph{\bibinfo{booktitle}{Star Clusters: From the Milky Way to the Early Universe}}, vol. \bibinfo{volume}{351}, \bibinfo{pages}{161--164} (\bibinfo{year}{2020}).
\newblock \eprint{1908.07278}.

\bibitem{Penarrubia-2016}
\bibinfo{author}{{Pe{\~n}arrubia}, J.}, \bibinfo{author}{{G{\'o}mez}, F.~A.}, \bibinfo{author}{{Besla}, G.}, \bibinfo{author}{{Erkal}, D.} \& \bibinfo{author}{{Ma}, Y.-Z.}
\newblock \bibinfo{title}{{A timing constraint on the (total) mass of the Large Magellanic Cloud}}.
\newblock \emph{\bibinfo{journal}{\mnras}} \textbf{\bibinfo{volume}{456}}, \bibinfo{pages}{L54--L58} (\bibinfo{year}{2016}).
\newblock \eprint{1507.03594}.

\bibitem{Sawala-2023a}
\bibinfo{author}{{Sawala}, T.}, \bibinfo{author}{{Teeriaho}, M.} \& \bibinfo{author}{{Johansson}, P.~H.}
\newblock \bibinfo{title}{{The Local Group's mass: probably no more than the sum of its parts}}.
\newblock \emph{\bibinfo{journal}{\mnras}} \textbf{\bibinfo{volume}{521}}, \bibinfo{pages}{4863--4877} (\bibinfo{year}{2023}).
\newblock \eprint{2210.07250}.

\bibitem{Sawala-2023}
\bibinfo{author}{{Sawala}, T.}, \bibinfo{author}{{Pe{\~n}arrubia}, J.}, \bibinfo{author}{{Liao}, S.} \& \bibinfo{author}{{Johansson}, P.~H.}
\newblock \bibinfo{title}{{The timeless timing argument and the total mass of the Local Group}}.
\newblock \emph{\bibinfo{journal}{\mnras}} \textbf{\bibinfo{volume}{526}}, \bibinfo{pages}{L77--L82} (\bibinfo{year}{2023}).
\newblock \eprint{2307.13732}.

\bibitem{DSouza-2018}
\bibinfo{author}{{D'Souza}, R.} \& \bibinfo{author}{{Bell}, E.~F.}
\newblock \bibinfo{title}{{The Andromeda galaxy's most important merger about 2 billion years ago as M32's likely progenitor}}.
\newblock \emph{\bibinfo{journal}{Nature Astronomy}} \textbf{\bibinfo{volume}{2}}, \bibinfo{pages}{737--743} (\bibinfo{year}{2018}).
\newblock \eprint{1807.08819}.

\bibitem{Libeskind-2020}
\bibinfo{author}{{Libeskind}, N.~I.} \emph{et~al.}
\newblock \bibinfo{title}{{The HESTIA project: simulations of the Local Group}}.
\newblock \emph{\bibinfo{journal}{\mnras}} \textbf{\bibinfo{volume}{498}}, \bibinfo{pages}{2968--2983} (\bibinfo{year}{2020}).
\newblock \eprint{2008.04926}.

\bibitem{Sawala-2022}
\bibinfo{author}{{Sawala}, T.} \emph{et~al.}
\newblock \bibinfo{title}{{The SIBELIUS Project: E Pluribus Unum}}.
\newblock \emph{\bibinfo{journal}{\mnras}} \textbf{\bibinfo{volume}{509}}, \bibinfo{pages}{1432--1446} (\bibinfo{year}{2022}).
\newblock \eprint{2103.12073}.

\bibitem{Wempe-2024}
\bibinfo{author}{{Wempe}, E.} \emph{et~al.}
\newblock \bibinfo{title}{{Constrained cosmological simulations of the Local Group using Bayesian hierarchical field-level inference}}.
\newblock \emph{\bibinfo{journal}{arXiv e-prints}} \bibinfo{pages}{arXiv:2406.02228} (\bibinfo{year}{2024}).
\newblock \eprint{2406.02228}.

\bibitem{Astropy}
\bibinfo{author}{{Astropy Collaboration}} \emph{et~al.}
\newblock \bibinfo{title}{{The Astropy Project: Sustaining and Growing a Community-oriented Open-source Project and the Latest Major Release (v5.0) of the Core Package}}.
\newblock \emph{\bibinfo{journal}{\apj}} \textbf{\bibinfo{volume}{935}}, \bibinfo{pages}{167} (\bibinfo{year}{2022}).
\newblock \eprint{2206.14220}.

\bibitem{numpy-paper}
\bibinfo{author}{Harris, C.~R.} \emph{et~al.}
\newblock \bibinfo{title}{Array programming with {NumPy}}.
\newblock \emph{\bibinfo{journal}{Nature}} \textbf{\bibinfo{volume}{585}}, \bibinfo{pages}{357--362} (\bibinfo{year}{2020}).
\newblock \urlprefix\url{https://doi.org/10.1038/s41586-020-2649-2}.

\bibitem{SciPy}
\bibinfo{author}{Virtanen, P.} \emph{et~al.}
\newblock \bibinfo{title}{{{SciPy} 1.0: Fundamental Algorithms for Scientific Computing in Python}}.
\newblock \emph{\bibinfo{journal}{Nature Methods}} \textbf{\bibinfo{volume}{17}}, \bibinfo{pages}{261--272} (\bibinfo{year}{2020}).

\bibitem{Diemer-2018}
\bibinfo{author}{{Diemer}, B.}
\newblock \bibinfo{title}{{COLOSSUS: A Python Toolkit for Cosmology, Large-scale Structure, and Dark Matter Halos}}.
\newblock \emph{\bibinfo{journal}{\apjs}} \textbf{\bibinfo{volume}{239}}, \bibinfo{pages}{35} (\bibinfo{year}{2018}).
\newblock \eprint{1712.04512}.

\end{thebibliography}
\clearpage


\newpage

\clearpage

\begin{figure*}
\centering
\vspace{-.4cm}
    \includegraphics[height=5.5cm, trim={0cm 0.2cm 0cm 0.6cm},clip]{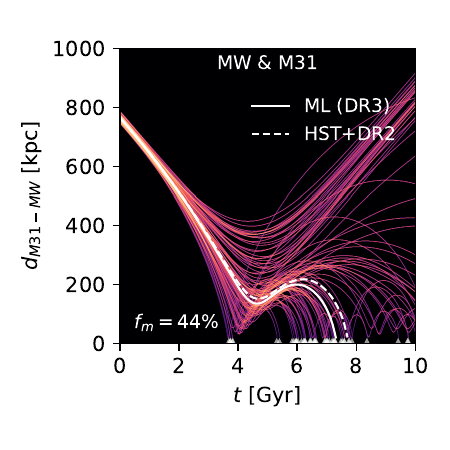}
    \includegraphics[height=5.5cm, trim={0cm 0.2cm 0cm 0.6cm},clip]{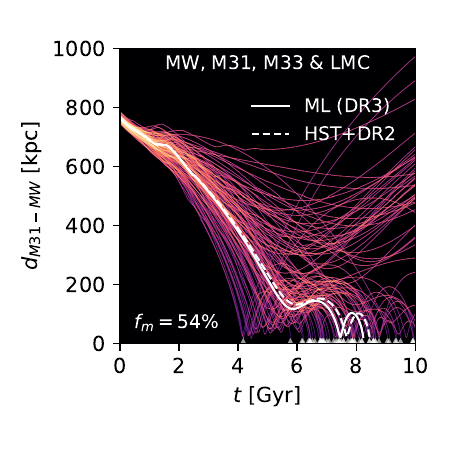} \\
    \includegraphics[height=5.5cm, trim={0cm 0.2cm 0cm 0.6cm},clip]{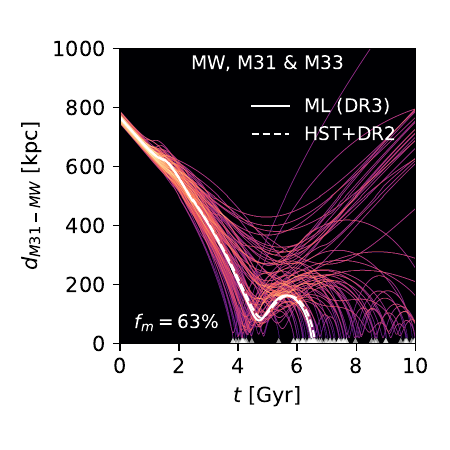}
    \includegraphics[height=5.5cm, trim={0cm 0.2cm 0cm 0.6cm},clip]{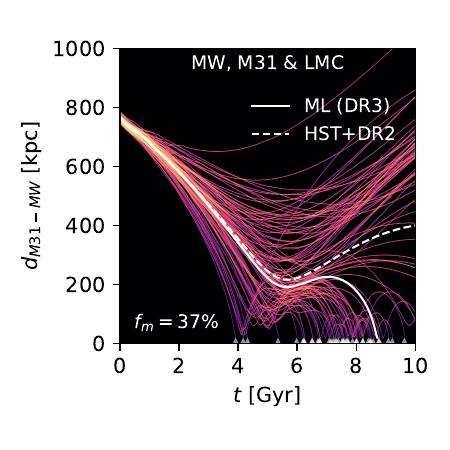}
    \caption{Distance between the MW and M31. On each panel, we show 100 realisations of our fiducial MC model and also state the probability for a MW-M31 merger within 10 Gyr. In the top row, we show the MW-M31 two-body orbits (top-left) and MW-M31-M33-LMC four-body orbits (top-right) shown in Figure~\ref{fig:orbits}. In the bottom row, we show the MW-M31-M33 (bottom-left) and MW-M31-LMC (bottom-right) three-body orbits. White markers denote MW-M31 mergers; percentages indicate the fraction of orbits that merge within 10 Gyr. Only slightly more than half of the four-body orbits lead to a merger within 10 Gyr. The inclusion of M33 increases the likelihood of a merger, whereas the inclusion of the LMC decreases it. White lines show the individual orbits using the most likely values of every variable, either assuming the Gaia DR3 proper motions \cite{Salomon-2021} of the fiducial model (solid) or HST + Gaia DR2 proper motions \cite{VanderMarel-2019} (dashed).}
    \label{fig:distances}
\end{figure*}
\newpage

\clearpage

\clearpage

\clearpage

\clearpage

\backmatter

\clearpage

\renewcommand\tablename{Extended Data Table}
\setcounter{figure}{0} 
\renewcommand\figurename{Extended Data Figure}
\section*{Methods}
Our results are based on numerically integrated initial conditions, which are in turn based on Monte Carlo samples of the observational data. We describe here the generation of the Monte Carlo samples, the numerical integration, and our treatment of dynamical friction and galaxy mergers. We also demonstrate the robustness of our findings to the particular choices made and show that our fiducial model and the results are conservative in predicting an {\it uncertain} future and relatively {\it low} merger probability. We also present results when the SMC is included in the analysis, in addition to the four main galaxies. To facilitate the reproduction of our results and allow future work to easily incorporate new observational data, our analysis code is flexible and public.

\subsection*{Monte Carlo Samples}
Our fiducial model consists of the Milky Way and three additional galaxies, M31, M33, and the LMC. We assume that the sky coordinates (RA \& Dec) for the centres of M31, M33, and the LMC are known. We furthermore assume that the position of the Galactic centre with respect to the sun is fixed at $(Ra, Dec)$ = (266.4051, -28.936175) \cite{Reid-2004}, $d=8.122$~kpc \cite{Gravity-2018}, and that the galactocentric velocity of the Sun is $(12.9, 245.6, 7.78)$~km$^{-1}$ \cite{Drimmel-2018, Gravity-2018, Reid-2004}.

We create Monte Carlo samples for the remaining 20 variables, the four halo masses, $M_{MW}$, $M_{M31}$, $M_{M33}$, $M_{LMC}$, the four halo concentration parameters, the three distance moduli, $\mu$, the three sets of proper motions, $\mu_\delta$ and $\mu_\alpha^*$, and three line-of-sight velocities, $v_\mathrm{los}$. Sampling directly in the space of the observables rather than sampling in a space of derived variables such as Cartesian coordinates or velocities minimises the effect of possible correlations.

To allow reproducibility and identification of individual orbits across our figures when changing parameters of the model such as the set of galaxies included and numerical parameters such as the gravitational softening length and merger threshold, we re-initialise the pseudo-random number generator with the same values for every new set of Monte Carlo samples. While we generally only show the first 100 orbits on each plot, all quoted probabilities are computed from at least 2500 samples each, so that statistical errors are less than $1\%$.

The most likely values of each variable together with the $\pm 1 \sigma$ uncertainty are either taken directly from single sources in the literature or in the case of masses and concentrations (see below), estimated by us based on multiple sources. In creating our Monte Carlo samples, we assume that each variable follows a Gaussian probability distribution. However, in our fiducial model, we truncate all distributions at $\pm 2\sigma$, corresponding to the central $\sim95\%$ of values. While Gaussian distributions are a natural assumption for measurement errors, this is not always explicitly stated, and it may not reflect the true probability distribution, especially at some distance from the most likely values. Indeed, for some variables, untruncated distributions extend to unphysical values with finite probabilities. A truncation at $2 \sigma$ also ensures that all variables in the samples are physical.

\begin{figure*}
\centering
\vspace{-.4cm}
    \includegraphics[height=4.5cm, trim={0cm 0.2cm 0cm 0.6cm},clip]{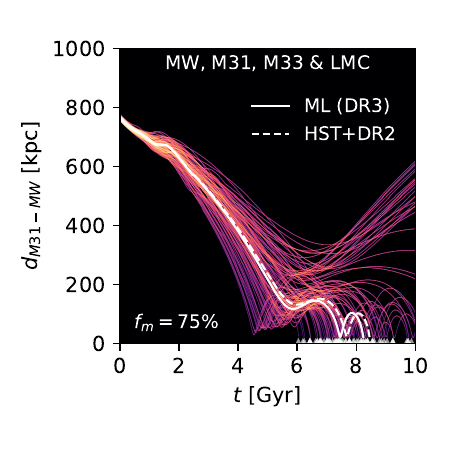}
    \includegraphics[height=4.5cm, trim={1.8cm 0.2cm 0cm 0.6cm},clip]{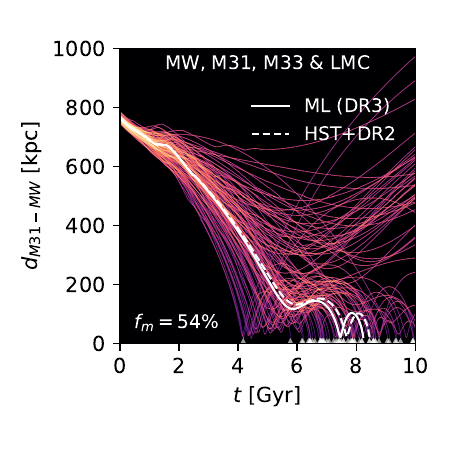}
    \includegraphics[height=4.5cm, trim={1.8cm 0.2cm 0cm 0.6cm},clip]{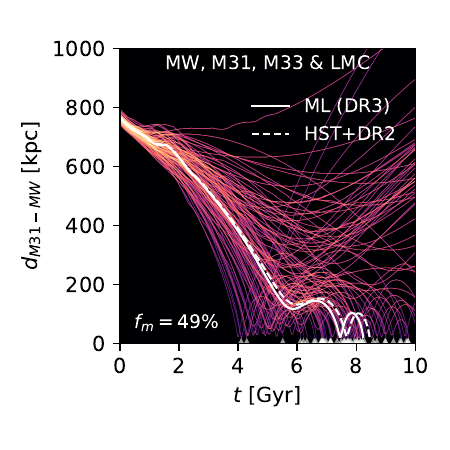}
    \caption{Effect of truncating the assumed probability distributions of observables. The distance between the MW and M31 (analogous to Figure~\ref{fig:distances}) in the four-body MW-M31-M33-LMC system for different truncations of the observables. From left to right, we show results where the probability distribution for each observable in the fiducial model is truncated to $\pm 1 \sigma$,  $\pm 2 \sigma$ (our default model, same as in Figure~\ref{fig:distances}), or left untruncated. The fraction of systems that merge is only slightly increased when the distributions are clipped at $\pm 2 \sigma$ or above.} 
    \label{fig:ext-truncation}
\end{figure*}

Given that our central claim is uncertainty in the future evolution of the LG, truncating the probability distributions of the observables is a conservative assumption that leads to lower uncertainty about the outcome. However, as shown in Extended~Data~Figure~\ref{fig:ext-truncation}, our results are not sensitive to the truncation at $\pm 2 \sigma$, and the merger probability is only slightly lower when the distributions are not truncated. On the other hand, even a truncation at only $\pm 1 \sigma$ leaves approximately $1/4$ of orbits that do not merge within 10 Gyr.

\subsection*{Numerical Integration}
The orbits are integrated using a symplectic leapfrog algorithm in the centre-of-mass frame, with a step size of 1 Myr, approximately the time it takes for a galaxy moving at 100 kms$^{-1}$ to travel 0.1 kpc --- 1/200$^{\rm th}$ of our merger threshold (see below). Our results are not affected by the finite time step.

\begin{figure*}
\centering
\vspace{-.4cm}

    \includegraphics[height=4.cm, trim={0cm 0.8cm 0cm 0.6cm},clip]{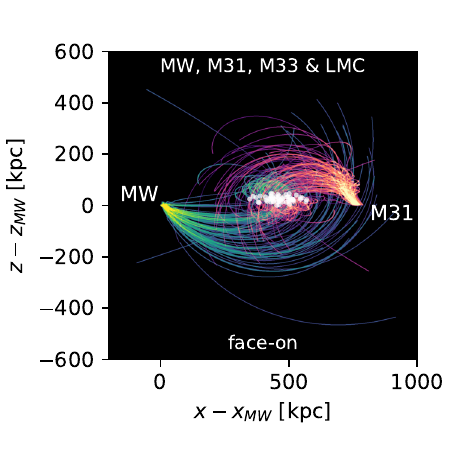}
    \includegraphics[height=4.cm, trim={1.6cm 0.8cm 0cm 0.6cm},clip]{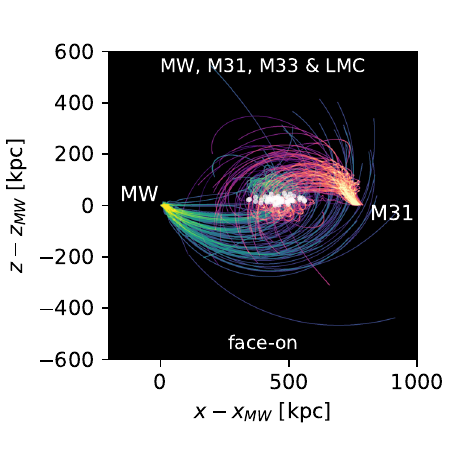} 
    \includegraphics[height=4.cm, trim={1.6cm 0.8cm 0cm 0.6cm},clip]{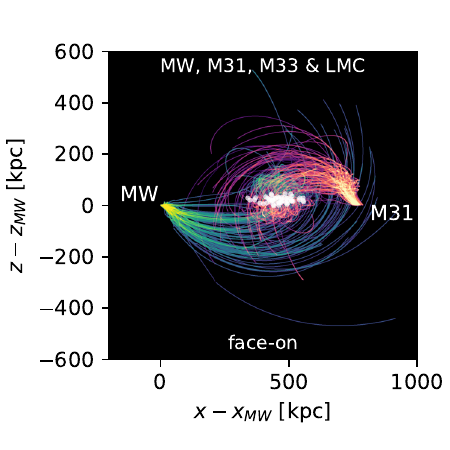}  \\

    \includegraphics[height=4.5cm, trim={0cm 0cm 0cm 0.6cm},clip]{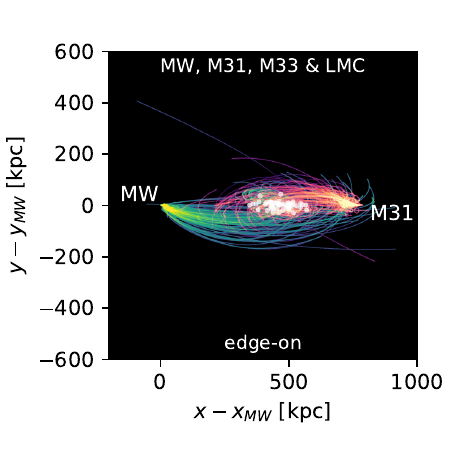}
    \includegraphics[height=4.5cm, trim={1.6cm 0cm 0cm 0.6cm},clip]{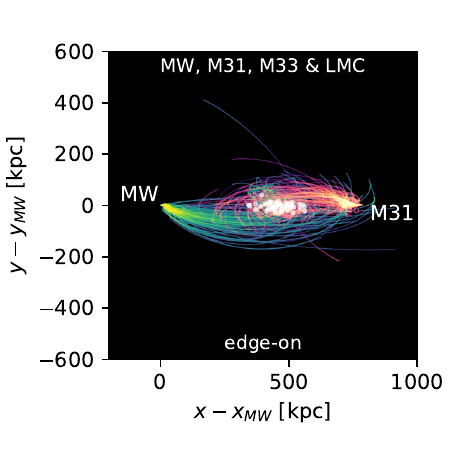} 
    \includegraphics[height=4.5cm, trim={1.6cm 0cm 0cm 0.6cm},clip]{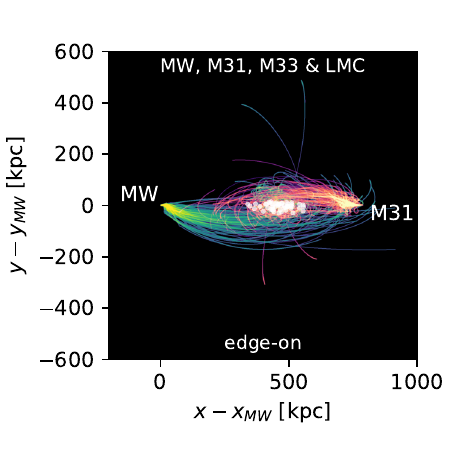} \\

    \includegraphics[height=4.55cm, trim={0.3cm 0.2cm 0cm 0.6cm},clip]{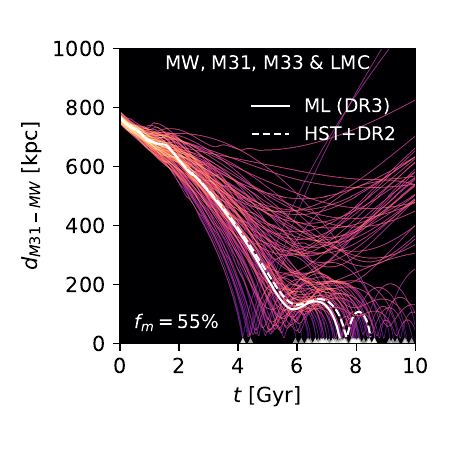}
    \includegraphics[height=4.55cm, trim={1.8cm 0.2cm 0cm 0.6cm},clip]{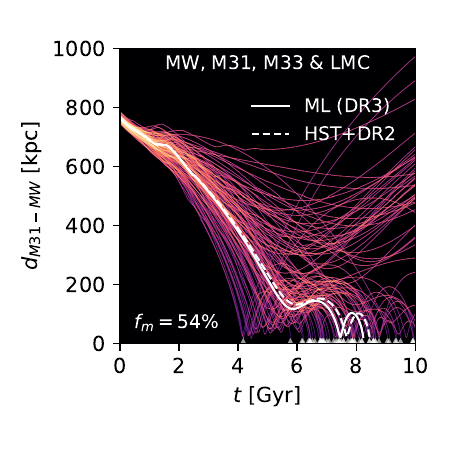}
    \includegraphics[height=4.55cm, trim={1.8cm 0.2cm 0cm 0.6cm},clip]{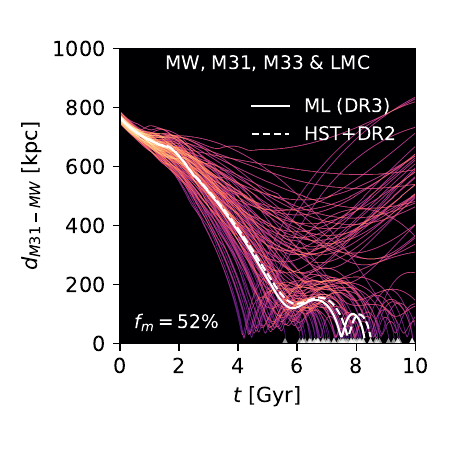}
    \caption{Effect of the gravitational softening. MW-M31 orbits (analogous to Figure~\ref{fig:orbits}) and distance between the MW and M31 (analogous to Figure~\ref{fig:distances}) in the four-body MW-M31-M33-LMC system for different softening lengths. From left to right, we show results with a softening length of 10~kpc, 20~kpc (our default value), and 30~kpc. A softening length that is too small can lead to some unrealistically strong kicks in close encounters, while a softening length that is too large weakens the overall gravitational attraction. However, the merger fraction is not significantly affected by the choice of softening length within this range.} 
    \label{fig:ext-softening}
\end{figure*}

To account for the fact that the haloes are extended objects, the gravitational force between the haloes is softened with a constant softening length of 20 kpc, similar to the scale radius of an NFW halo in the mass range we consider here. We also consider different choices for the softening and show in Extended Data Figure~\ref{fig:ext-softening} results with softening lengths of 10, 20, or 30 kpc, respectively. A softening length that is too small can lead to unphysical hard scattering events during close encounters while a softening length that is too large artificially reduces the gravitational force. In the context of the Local Group, both of these effects could reduce the merger probability. However, we find no strong dependence of the merger probability on the softening length, with our adopted fiducial value of 20 kpc resulting in the highest merger probability.

\subsection*{Dynamical Friction}
To estimate the effect of dynamical friction, we use a modified Chandrasekhar formula, similar to that used in \cite{Cautun-2019}. The classic Chandrasekhar formula assumes a point mass orbiting in the potential of a much more massive host halo, which is in turn composed of much less massive particles. This approach has been expanded to account for extended satellites \citeMethods{White-1976}, and we use the following expression to calculate the acceleration of a satellite due to dynamical friction \cite{Binney-Tremaine-2008} once inside $r_{200}$ of the host halo:

\begin{equation}
	\frac{{\rm d}{\bf v}}{{\rm d}t} = - \frac{ 4 \pi G^2 M \rho \ln\Lambda } {v^2} 
    	\left[ {\rm erf}(X) - \frac{2X}{\sqrt\pi} e^{-X^2} \right]\frac{{\bf v}}{v}
	\label{eq:chandrasekhar},
\end{equation}

where $G$ is the gravitational constant, $M$ is the mass of the satellite, ${\bf v}$ is the velocity of the satellite relative to the host, $\rho$ is the density of the host at the position of the satellite, $X = v / (2 \sigma)$ is the ratio between the orbital speed of the satellite and the 1D-velocity dispersion, $\sigma$, of the host at the location of the satellite, and $\Lambda$ is the Coulomb factor expressed as $r / \epsilon$, with $\epsilon$ a scale length that depends on the density of the satellite. To determine $\epsilon$, we adopt an empirical expression derived from N-body simulations \citeMethods{Jethwa-2016}:

\begin{equation}
\epsilon =
\begin{cases}
2.2 \ r_s - 14 ~\mathrm{kpc} & \text{if } r_s \geq 8 \mathrm{kpc} \\
0.45 \ r_s  & \text{if } r_s < 8 \mathrm{kpc}
\end{cases}
\label{eq:coulomb}
\end{equation}
where $r_s$ is the scale radius of the satellite's NFW halo. Finally, to approximate the velocity dispersion of the host at the location of the satellite, we use the expression derived in \cite{Zentner-2003} for NFW haloes.

Standard dynamical friction schemes assume a clear hierarchy between the (much more massive) host and the (much less massive) satellite. According to the assumptions underlying Equation~\ref{eq:chandrasekhar}, the satellite and host enter the calculation in clearly defined and distinct roles, with the dynamical friction force applied only on the satellite, while the host remains unaffected.

However, in the Local Group context where galaxies and haloes of similar mass are interacting, this introduces an inconsistency. In particular, in the (relatively likely) scenario that M31 has a similar mass to the MW, the roles of the satellite and host are unclear, but their assignment changes the result of the calculation. For example, if the MW is considered the satellite, it would be accelerated by its interaction with M31 while M31 would remain unaffected, and only the motion of the MW with respect to the other galaxies would be affected while that of M31 would remain unchanged. If M31 is considered the satellite, the roles would be reversed. Accelerating only one galaxy also violates momentum conservation.

To make the calculation more symmetrical, in our dynamical friction calculation we distribute the dynamical friction force proportionally between the satellite ($s$) and host ($h$), conserving the total momentum:

\begin{align} 
\frac{{\rm d}{\bf v}_s}{{\rm d}t}  &= {\bf a}_{DF} \ \frac{M_h}{M_s + M_h}, \\
\frac{{\rm d}{\bf v}_h}{{\rm d}t}  &= - {\bf a}_{DF} \ \frac{M_s}{M_s + M_h},
\end{align}
where ${\bf v}_s$ and $M_s$ are the velocity and mass of the satellite, ${\bf v}_h$ and $M_h$ are the mass and velocity of the host, and ${\bf a}_{DF}$ is the acceleration computed using Equation~\ref{eq:chandrasekhar}. In the limit that the satellite is much less massive than the host, the standard hierarchical scheme is recovered and only the satellite is accelerated, while in the limit that both galaxies have equal mass, both receive equal and opposite accelerations.

A small inconsistency remains in that even when the differences in mass are small, we still assign the more massive galaxy as the host and the less massive galaxy as the satellite when calculating the magnitude of the dynamical friction, where the velocity dispersion of the host, $\sigma$, but not that of the satellite, and through the Coulomb factor, the scale radius of the satellite, but not that of the host, are considered. In our spherical halo models, both the velocity dispersion and Coulomb factor depend only on the assumed masses and concentrations, and as we discuss below, the concentration of the satellite has a greater impact on the dynamical friction force. For two halos of significantly different masses, the assignment of host and satellite is clear. For an individual case of two haloes of nearly identical masses but different (randomly assigned) concentrations, the choice of calculating the dynamical friction force by treating either halo as the satellite seems arbitrary. However, for a large number of samples of nearly equal-mass interactions with randomly assigned concentrations, the dynamical friction calculations are not biased. We also assume identical distributions of concentration parameters for all galaxies.

\begin{figure*}
\centering

    \includegraphics[height=4.4cm, trim={0cm 0.2cm 0cm 0.6cm},clip]{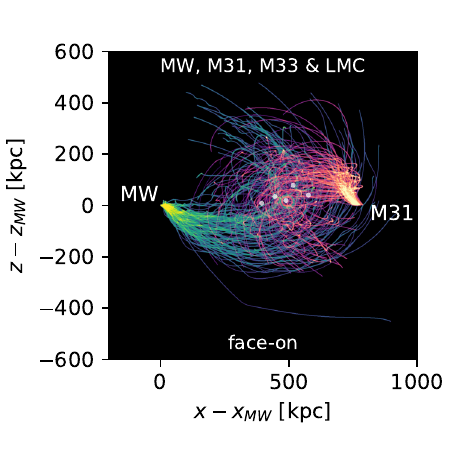}
    \includegraphics[height=4.4cm, trim={1.6cm 0.2cm 0cm 0.6cm},clip]{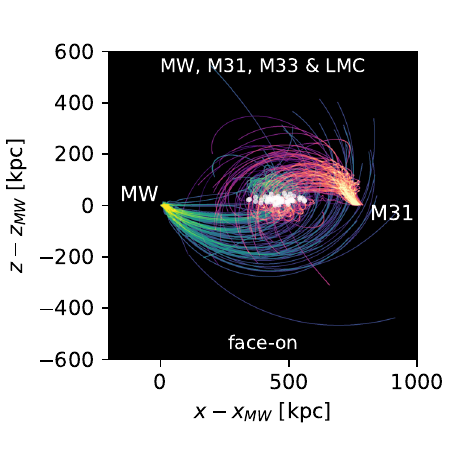}
    \includegraphics[height=4.4cm, trim={1.6cm 0.2cm 0cm 0.6cm},clip]{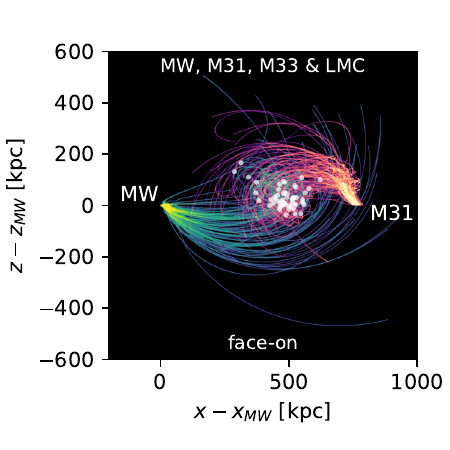} \\
    \includegraphics[height=4.4cm, trim={0cm 0.2cm 0cm 0.6cm},clip]{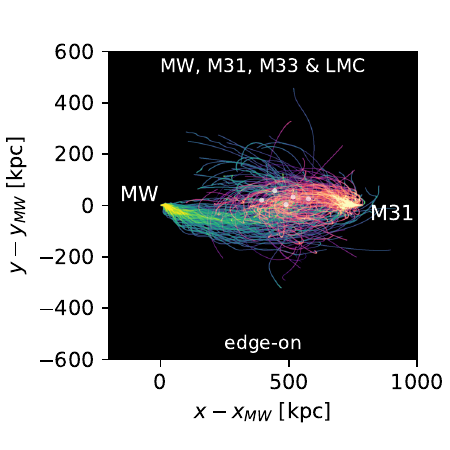}
    \includegraphics[height=4.4cm, trim={1.6cm 0.2cm 0cm 0.6cm},clip]{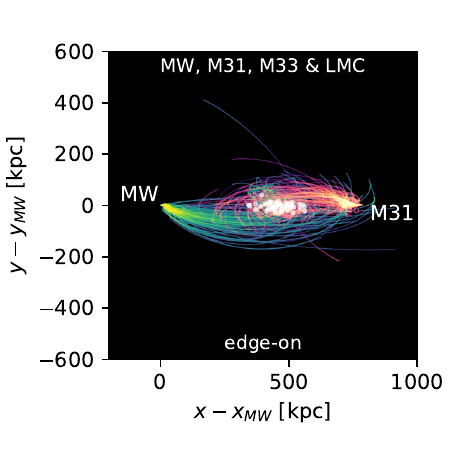}
    \includegraphics[height=4.4cm, trim={1.6cm 0.2cm 0cm 0.6cm},clip]{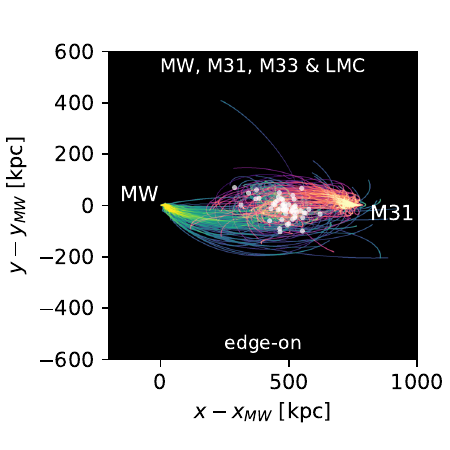} \\
    \includegraphics[height=4.5cm, trim={0cm 0.2cm 0cm 0.6cm},clip]{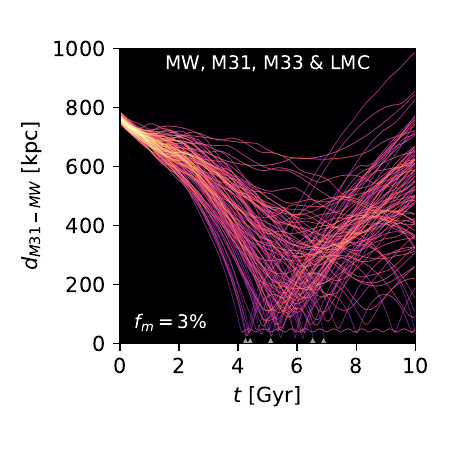}
    \includegraphics[height=4.5cm, trim={1.8cm 0.2cm 0cm 0.6cm},clip]{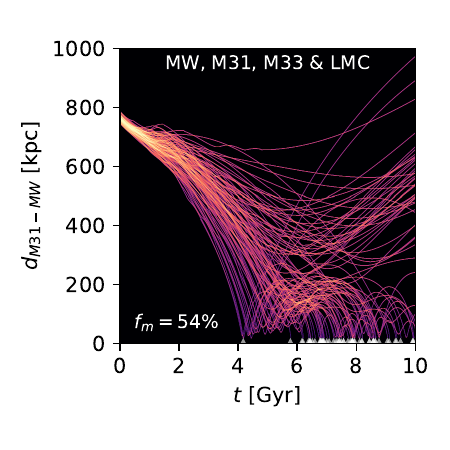}
    \includegraphics[height=4.5cm, trim={1.8cm 0.2cm 0cm 0.6cm},clip]{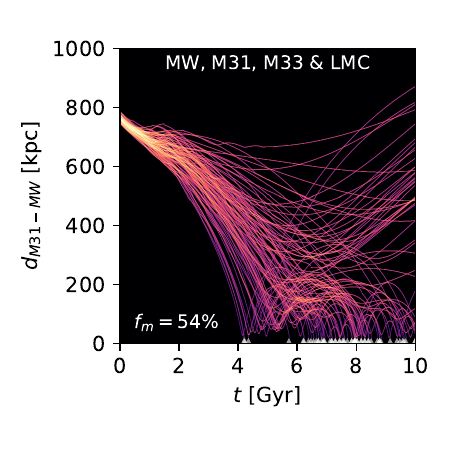}
    \caption{Effect of different schemes of dynamical friction. MW-M31 orbits (topt two rows, analogous to Figure~\ref{fig:orbits}) and distance between the MW and M31 (bottom row, analogous to Figure~\ref{fig:distances}) in the four-body MW-M31-M33-LMC system for different softening lengths. The left column assumes no dynamical friction, the middle column uses our default ``proportional" scheme where the dynamical friction force is divided such that equal and opposite dynamical forces are applied to both host and satellite, the right column uses the ``hierarchical" scheme where dynamical friction is only applied to the less massive object. Dynamical friction is essential for orbits to decay and for the MW-M31 merger to occur, but the probability of an MW-M31 merger is not sensitive to the scheme used. However, due to its momentum-conserving property, mergers in the proportional scheme are more likely to occur close to the original orbital plane compared to the hierarchical scheme.} 
    \label{fig:ext-dynamical-friction}
\end{figure*}

In Extended Data Figure~\ref{fig:ext-dynamical-friction}, we compare the orbits of the fiducial system using no dynamical friction, ``hierarchical'' dynamical friction (only from the more massive host to the less massive satellite), and our default ``proportional'' dynamical friction. It is clear that without dynamical friction a MW-M31 merger is highly unlikely. In fact, the finite merger rate without dynamical friction depends strongly on our default impact parameter threshold of 20 kpc, with a lower threshold, the merger rate can become arbitrarily small. On the other hand, when dynamical friction is included, the evolution of each orbit is quite similar in the ``hierarchical'' and ``proportional'' schemes, and the merger rate is not significantly affected by the exact choice of scheme.

It is worth noting that our semi-analytical approach to dynamical friction is still quite simplistic, and while the average behaviour of N-body simulations has been used to calibrate parameters, numerical simulations also show that individual systems with non-zero internal angular momenta and substructures can have different merger times than predicted by these simple equations \citeMethods{Boylan-Kolchin-2008}. A precise prediction of the MW-M31 orbit will likely require full N-body simulations. On the other hand, we show that even a simple dynamical model that assumes no spin, no triaxiality, and no substructure results in considerable uncertainty in the future evolution of the MW-M31 system.

\subsection*{Mergers}
Below a certain distance, the interactions of the gas and stellar components become significant, and our simple approach is no longer appropriate for predicting the remaining orbital evolution. Our aim here is not to predict the precise time of the merger (in fact, we argue that such a prediction is futile based on the current data), so we simply assume that any system that passes below a threshold distance will eventually merge, and identify this time as a lower limit for the time of the merger.

In our fiducial model, we have adopted a value of 20 kpc as the merger threshold for all galaxy interactions. When such a merger occurs between two galaxies, we combine the masses and momenta of the two galaxies at their common centre of mass and continue the integration. The concentration parameter is set to that of the most massive galaxy.

\begin{figure*}
\centering
\vspace{-.4cm}

    \includegraphics[height=4.5cm, trim={0cm 0.2cm 0cm 0.6cm},clip]{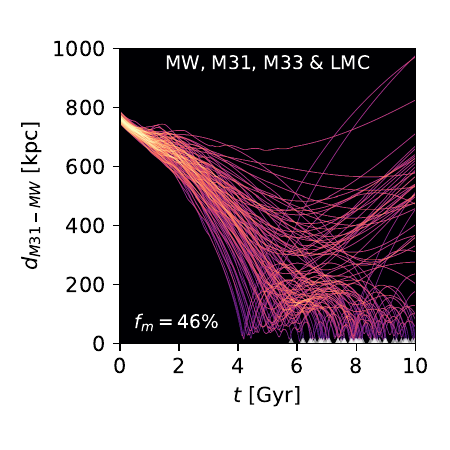}
    \includegraphics[height=4.5cm, trim={1.8cm 0.2cm 0cm 0.6cm},clip]{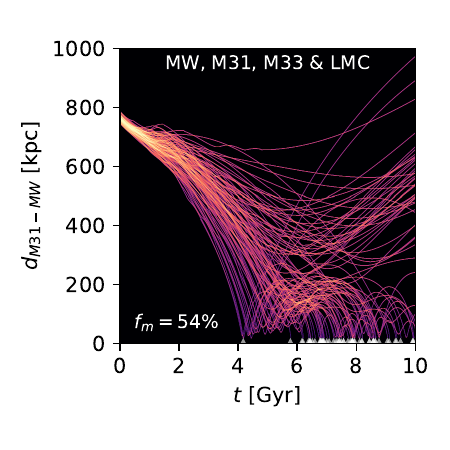}
    \includegraphics[height=4.5cm, trim={1.8cm 0.2cm 0cm 0.6cm},clip]{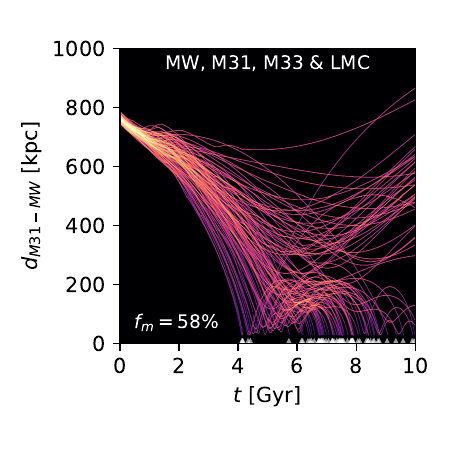}
    \caption{Effect of the merger threshold on the MW-M31 distance evolution and merger rate. The distance between the MW and M31 (analogous to Figure~\ref{fig:distances}) in the four-body MW-M31-M33-LMC system for different merger thresholds. From left to right, we show results with a threshold of 10~kpc, 20~kpc or 30~kpc for all mergers. The MW-M31 merger probability is not very sensitive to the assumed merger threshold, and we obtain almost the same merger probability even with a (very generous) threshold of 30 kpc.} 
    \label{fig:ext-threshold}
\end{figure*}

In Extended Data Figure~\ref{fig:ext-threshold}, we compare the sensitivity of our results to adopting merger thresholds of 10, 20, and 30~kpc. With a threshold of 10~kpc, we find a slightly reduced MW-M31 merger probability. However, raising the threshold from 20 to 30 kpc has no significant impact on our results. This confirms our assertion in the main section: due to the effect of dynamical friction, orbits either inspiral and eventually merge, or do not come close enough for dynamical friction to become effective and hence do not merge.

The small fraction of MW-M31 orbits that merge with a threshold of 20 kpc, but do not when the threshold is set to 10 kpc, approach with a small impact parameter and high velocity. This reduces the effect of dynamical friction and also allows them to escape to a large apocentre. While this may be a possible scenario for the Local Group, our methods are not adequate for studying such close interactions between galaxies. To be conservative in our prediction of a low merger probability, by setting the merger threshold at 20 kpc we assume that these orbits also merge, and with an even larger threshold of 30 kpc, we would still predict a similar merger rate.

For comparison, the best-studied merger of two galaxies with similar stellar masses to the MW and M31 are the Antenna galaxies. Their past evolution is reproduced with an orbit with pericentre $\sim 10$~kpc \citeMethods{Karl-2010}, predicted to lead to coalescence $\sim 1.3$~Gyr later \citeMethods{Lahen-2018}.

\subsection*{Fate of the LMC and M33}
While our main focus here is on the future evolution of the MW-M31 orbit, we naturally also make predictions for the evolution of the LMC and M33.

In the fiducial model, we find that the LMC is certain to merge with the MW before any eventual MW-M31 merger. With a merger threshold of 20 kpc, we find a median time of the LMC-MW merger of 1.3 Gyr, while with a threshold of 10 kpc, we find a median time of 1.9 Gyr.

For M33, with a merger threshold of 20~kpc, we find an $\sim 86\%$ chance of a merger with M31 and a median time of 3.3 Gyr, while with a merger threshold of 10~kpc, we find an $\sim 83 \%$ probability of a merger with M31 and a median time of 3.9 Gyr. In both cases, we also find a small probability of $\sim 1-2 \%$ for a merger of M33 with the MW-M31 remnant after a MW-M31 merger within the next 10 Gyr.

It must be noted that our simulations are not designed to study these mergers in detail and ignore, for example, the impact of the disk of the Milky Way. Nevertheless, they broadly agree with the results of \cite{Cautun-2019}, who have previously studied the LMC-MW encounter in the presence of M31.

\subsection*{Additional galaxies}
\begin{figure*}
\centering
\vspace{-.4cm}
    \includegraphics[height=5.5cm, trim={0cm 0.2cm 0cm 0.6cm},clip]{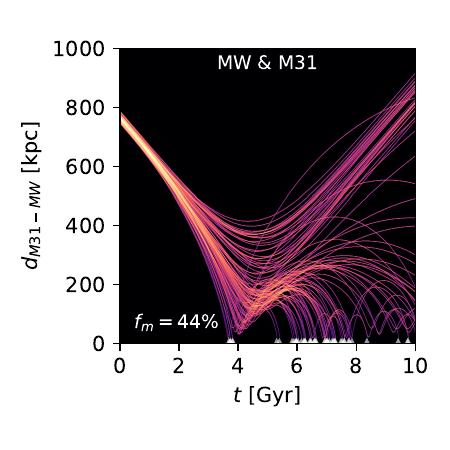}
    \includegraphics[height=5.5cm, trim={0cm 0.2cm 0cm 0.6cm},clip]{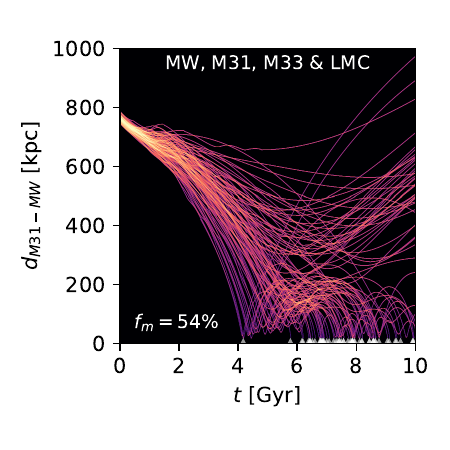}
    \includegraphics[height=5.5cm, trim={0cm 0.2cm 0cm 0.6cm},clip]{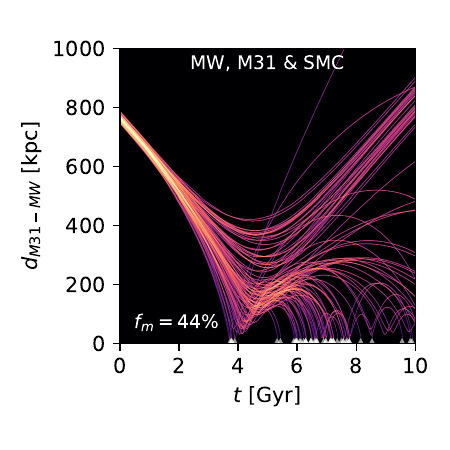}
    \includegraphics[height=5.5cm, trim={0cm 0.2cm 0cm 0.6cm},clip]{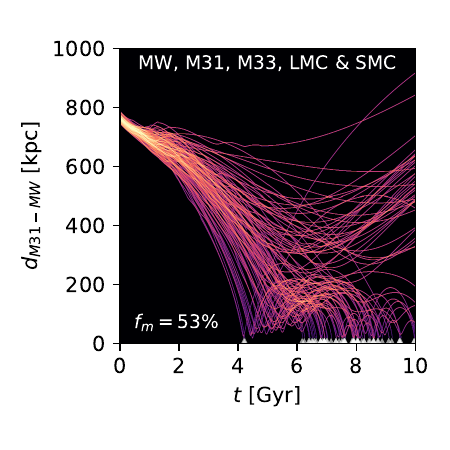}
    \caption{The effect of including the SMC on the MW-M31 distance and merger rate. In the top row, we show the MW-M31 two-body system, and the MW-M31-M33-LMC four-body system in the fiducial model, corresponding to the top row of Figure~\ref{fig:distances}. In the bottom row, we add the SMC to both systems, i.e. we show the MW-M31-SMC three-body system and the MW-M31-M33-LMC-SMC five-body system. The inclusion of the SMC has only a small effect on most MW-M31 orbits and does not significantly affect the total merger rate.} 
    \label{fig:ext-SMC}
\end{figure*}

The next most massive LG member galaxy for which proper motion data \cite{Kallivayalil-2013} is available is the Small Magellanic Cloud (SMC), a satellite galaxy of the MW that has likely been accreted together with the LMC. We repeat our analysis including the SMC as a fifth system, and show results in Extended Data Figure~\ref{fig:ext-SMC}. Adding the SMC, whose mass is approximately $10\%$ of that of the LMC, has no significant effect on the merger rate. The SMC properties used are listed along with those of the other galaxies in Extended Data Table~\ref{tab:data}.

\subsection*{DR2 + HST proper motions}
\begin{figure*}
\centering
\vspace{-.4cm}
    \includegraphics[height=5.5cm, trim={0cm 0.2cm 0cm 0.6cm},clip]{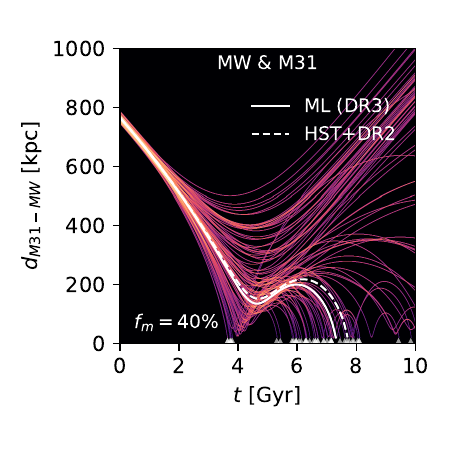}
    \includegraphics[height=5.5cm, trim={0cm 0.2cm 0cm 0.6cm},clip]{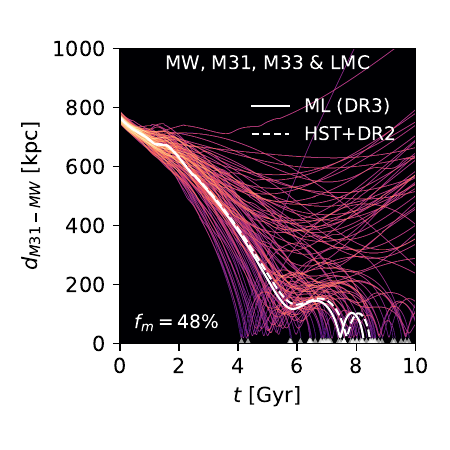} \\
    \includegraphics[height=5.5cm, trim={0cm 0.2cm 0cm 0.6cm},clip]{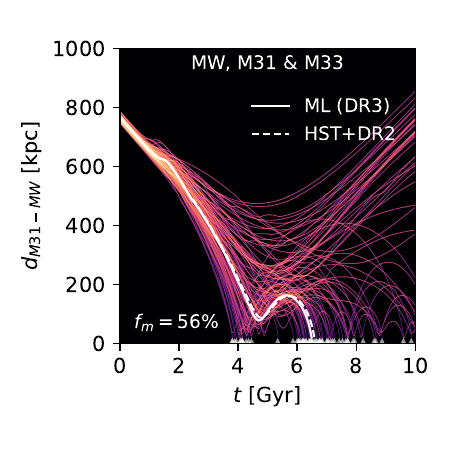}
    \includegraphics[height=5.5cm, trim={0cm 0.2cm 0cm 0.6cm},clip]{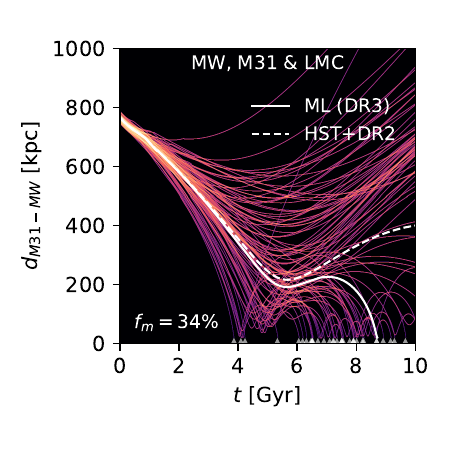}
    \caption{Distance between the MW and M31, using  HST+\textit{Gaia} DR2 proper motions for M31 \cite{VanderMarel-2019}, analogous to Figure~\ref{fig:distances}. As in Figure~\ref{fig:distances}, solid and dashed white lines denote the orbits using the most likely values using either \textit{Gaia}~DR3 proper motions \cite{Salomon-2021} or HST+\textit{Gaia}~DR2 proper motions. White markers denote MW-M31 mergers, percentages indicate the fraction of orbits that merge within 10 Gyr. The merger rate is slightly lower in all cases when compared to the fiducual model that uses the more precise \textit{Gaia}~DR3 proper motions, but the results are qualitatively similar.} 
    \label{fig:ext-distances-vDM}
\end{figure*}

Due to the fact that the M31 proper motions have the largest impact on the probability of a MW-M31 orbit, and for easier comparison with earlier works, particularly \cite{VanderMarel-2019}, we also repeat our analysis adopting the HST+{\it Gaia DR2} M31 proper motions of \cite{VanderMarel-2019}. In Extended Data Figure~\ref{fig:ext-distances-vDM}, we show the corresponding evolution of the MW-M31 distance in the same two-body MW-M31, three-body MW-M31-M33 and MW-M31-LMC, and four-body MW-M31-M33-LMC systems. The results can be directly compared to Figure~\ref{fig:distances} which shows the same quantities in our fiducial model that used {\it Gaia} DR3 proper motions. In each case, the merger probability is slightly lower using the HST+{\it Gaia DR2} proper motions: $40\%$ instead of $44\%$ for the MW-M31 system,  $56\%$ instead of $63\%$ for the MW-M31-M33 system,  $34\%$ instead of $37\%$ for the MW-M31-LMC system, and $48\%$ instead of $54\%$ for the full MW-M31-M33-LMC system. However, both sources of proper motions predict similar distributions of outcomes and a similar uncertainty about the MW-M31 merger. This difference can be attributed to the slightly lower precision of the HST+{\it Gaia DR2} proper motions.

\subsection*{Sources for the masses}
As discussed in Figure~\ref{fig:distributions}, the assumed masses of all four galaxies and their associated uncertainties have a strong impact on the likely evolution of the MW-M31 orbit in our fiducial model. Here, we review the mass measurements and show the implications for some alternative scenarios.

{\bf MW} \ \ The total mass of the MW has been extensively studied with different methods and tracers, and the accurate astrometry of the \textit{Gaia} space telescope has brought a flurry of recent measurements. Estimates for the total MW mass based on \textit{Gaia} DR2 or DR3 satellite dynamics include $1.17_{-0.15}^{+0.21} \times 10^{12} \Ms$ \citeMethods{Callingham-2019}, $1.51^{0.45}_{-0.40}\times 10^{12} \Ms$ \citeMethods{Fritz-2020}, $1.23^{+0.21}_{-0.18} \times 10^{12} \Ms$ \citeMethods{Li-2020}, and $1.1_{-0.1}^{+0.1} \times 10^{12} \Ms$ \citeMethods{Rodriguez-Wimberly-2022} (where the latter two works also use a simulation based prior). 

Estimates using rotation curves include $1.08^{+0.20}_{-0.14} \times 10^{12} \Ms$ based on Gaia DR2 \citeMethods{Cautun-2020}, $0.89^{+0.1}_{-0.08} \times 10^{12} \Ms$ using stars in the {\it galkin} catalogue \citeMethods{Karukes-2020}, $0.822 \pm 0.052 \times 10^{12} \Ms$ using classical Cepheids \citeMethods{Ablimit-2020}, and $1.08^{+0.12}_{-0.11} \times 10^{12} \Ms$ from the H3 survey and \textit{Gaia} DR3 \citeMethods{Shen-2022}.  Other recent measurements include  $1.54_{-0.44}^{+0.75} \times 10^{12} \Ms$ using combined \textit{Gaia} DR2 and HST kinematics of globular clusters \citeMethods{Watkins-2019}, $1.16 \pm 0.24 \times 10^{12} \Ms$ (including the mass of the LMC) using the kinematics of halo stars, $1.26^{+0.40}_{-0.22} \times 10^{12} \Ms$ from high velocity RR-Lyrae stars \citeMethods{Prudil-2022}, and $1.19_{-0.32}^{+0.49} \times 10^{12} \Ms$ from a Bayesian estimate using dwarf galaxy kinematics from multiple sources \citeMethods{Slizewski-2022}. 

\cite{Sawala-2023a} contains an overview of recent measurements and \citeMethods{Wang-2020} includes a comprehensive review of earlier results.

Analysis from several different tracers, and particularly from the latest studies using the latest \textit{Gaia} observations consistently point towards a Milky Way total mass that is close to $10^{12} \Ms$. Both the simple mean and median of the above measurements are $1.16 \times 10^{12} \Ms$, but it is worth noting that most methods of measuring the total mass of the MW include that of the LMC (at a galactocentric distance of $\sim 50$~kpc, well within $r_{\rm vir}$ or $r_{200}$), which is not always made explicit. When the LMC is excluded, the mass of the MW is reduced by $\sim 0.15 \times 10^{12} \Ms$ (see our discussion of the LMC mass below). We adopt $M_{MW} = 1.0 \pm 0.2 \times 10^{12} \Ms$ excluding the LMC in our fiducial model, reflecting the consensus of recent observations.

{\bf M31} While M31 does not enjoy the benefit of accurate \textit{Gaia} proper motions, there nevertheless exist a number of studies estimating its mass. Most recently, \citeMethods{Zhang-2024} measured a total mass of $1.14^{+0.51}_{-0.35} \times 10^{12} \Ms$ using rotation curves based on LAMOST data release 9 and DESI. \citeMethods{Watkins-2010} measured  $1.4 \pm 0.4 \times 10^{12} \Ms$ derived from satellite kinematics, \citeMethods{Tollerud-2012} derived $1.2_{-0.7}^{+0.9} \times 10^{12} \Ms$ from kinematics of M31 dwarf spheroidals, and \citeMethods{Tamm-2012} measured a total mass of $1.05^{+0.15}_{-0.15} \times 10^{12} \Ms$ from SED fitting together with the rotation curve and the kinematics of outer globular clusters and satellite galaxies. \citeMethods{Chemin-2009} measured $1.0 \times 10^{12} \Ms$ from the HI rotation curve, \citeMethods{Fardal-2013} measured $2.0^{+0.4}_{-0.3} \times 10^{12} \Ms$ from kinematics of the Giant southern stream. \citeMethods{Veljanoski-2013} measured $1.35_{-0.15}^{+0.15} \times 10^{12} \Ms$ and \citeMethods{Veljanoski-2014} found $1.4_{-0.2}^{+0.2} \times 10^{12} \Ms$ both using outer halo globular clusters, while \citeMethods{Kafle-2018} found $0.8 \pm 0.1 \times 10^{12} \Ms$ from high-velocity planetary nebulae, and \citeMethods{Sofue-2015} found $1.39 \pm 0.26 \times 10^{12}$ for the total mass combining disk rotation velocities and radial velocities of satellite galaxies and globular clusters. Both the simple mean and median of the above values are $1.27 \times 10^{12} \Ms$. Our adopted mass of $M_{M31} = 1.3\pm 0.4 \times 10^{12} \Ms$ in the fiducial model reflects the broad consensus of M31 mass estimates using different methods, but also the considerable remaining uncertainty.

{\bf M33} \ \ Mass estimates of M33 are much more sparse. More than twenty years ago, \citeMethods{Corbelli-2000, Corbelli-2003} measured a dark matter mass of $5 \times 10^{10} \Ms$, extrapolated out to a virial mass of $5\times 10^{11} \Ms$ from the measured HI rotation curve, but noted that this results in a very low baryon fraction. More recently, \citeMethods{Kam-2017} obtained a similar result using the $H\alpha$ rotation curve, but noted that because the measurements only extend to a few percent of the virial radius, there are no strong constraints on the total dark matter halo. On the other hand, abundance matching based on the observed stellar mass results in a significantly lower total mass of $1.7 \pm 0.55 \times 10^{11} \Ms$ \citeMethods{Guo-2010, Patel-2017}. Citing both the direct measurements and abundance matching, \citeMethods{Patel-2017} adopt a mass range of $0.8 - 3.2 \times 10^{11} \Ms$ in their dynamical models of the M33-M31 interaction, while \citeMethods{Patel-2018} and \cite{VanderMarel-2019} assume a total mass of $2.5 \times 10^{11} \Ms$.

We adopt $M_{M33} = 3 \pm 1 \times 10^{11} \Ms$ in the fiducial model, marginally compatible with both the extrapolated masses from rotation curve measurements and the results of abundance matching and in line with previous studies. However, the results from the two methods are certainly in tension.

{\bf LMC} \ \ For the mass of the LMC, recent mass estimates based on its effect on Galactic stellar streams include $1.38^{+0.37}_{-0.24} \times 10^{11} \Ms$ \citeMethods{Erkal-2019}, $1.30 \pm 0.3 \times 10^{11} \Ms$ \citeMethods{Vasiliev-2021}, $1.88^{+0.4}_{-0.35} \times 10^{11} \Ms$ \citeMethods{Shipp-2021} and $1.29^{+0.28}_{-0.23} \times 10^{11} \Ms$. \citeMethods{Erkal-2021} obtain good agreement between the perturbations of Milky Way halo stars with an LMC mass of $1.5 \times 10^{11} \Ms$. Using the abundance of likely satellites of the LMC, \citeMethods{Erkal-2020} obtain a lower limit of $1.24\times 10^{11} \Ms$, and using kinematics of satellites associated with the LMC, \citeMethods{Correa-Magnus-2022} find $1.65^{+0.47}_{-0.49} \times 10^{11} \Ms$. Most recently, \citeMethods{Watkins-2024} used 30 LMC globular clusters to infer a total mass of $1.80^{+1.05}_{-0.54}\times 10^{11} \Ms$ while \cite{Penarrubia-2016} find an even higher value of $2.5^{+0.9}_{-0.8} \times 10^{11} \Ms$ using a timing argument and to be most consistent with the Hubble Flow around the Local Group. The simple mean of the above values is $1.6 \times 10^{11} \Ms$ while the median is $1.5 \times 10^{11} \Ms$. A comprehensive recent review on the effect of the LMC on the Milky Way, including a discussion of mass measurements, is given by \citeMethods{Vasiliev-2023}, who conclude that the LMC mass is likely in the range $1-2 \times 10^{11} \Ms$, which matches the choice of $M_{LMC} = 1.5 \pm 0.5 \times 10^{11} \Ms$ in our fiducial model.

\subsection*{Halo Concentrations}
As explained above, we assume that for the purpose of integrating their orbits, all galaxies are represented by NFW haloes \cite{Navarro-1997} with the total masses $M_{200}$ defined above. We assume concentration parameters of $10 \pm 2$ for all galaxies, consistent with results of cosmological simulations \citeMethods{Neto-2007, Ludlow-2014, Wang-2015, Correa-2015, Diemer-2015} in this mass range. While cosmological simulations show the average concentration parameter to be mass-dependent, the halo-to-halo scatter is significantly larger than the change in the mean concentration over this narrow mass range \citeMethods{Neto-2007}. 

The concentration parameter does not affect the orbital calculation, except for the dynamical friction, where (for a given mass), it sets the scale radius, $r_s = r_{200} / c$, of the ``satellite" that enters in Equation~\ref{eq:coulomb}.

\begin{figure*}
\centering
\vspace{-.4cm}
    \includegraphics[height=4.5cm, trim={0cm 0.2cm 0cm 0.6cm},clip]{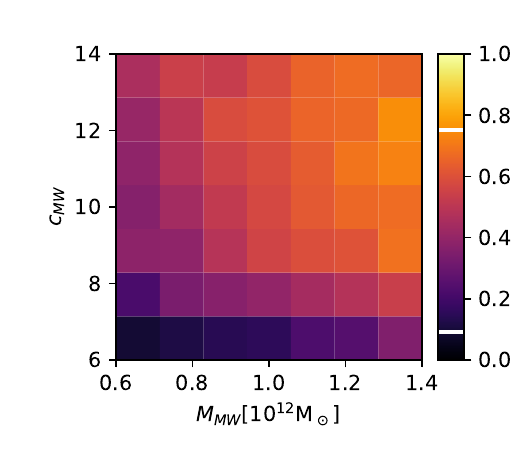}
    \includegraphics[height=4.5cm, trim={0cm 0.2cm 0cm 0.6cm},clip]{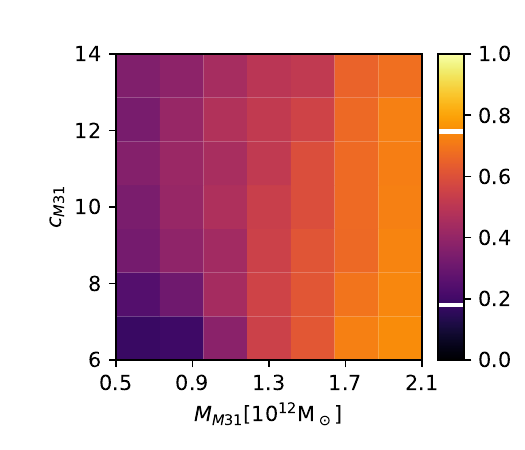} \\
    \includegraphics[height=4.5cm, trim={0cm 0.2cm 0cm 0.6cm},clip]{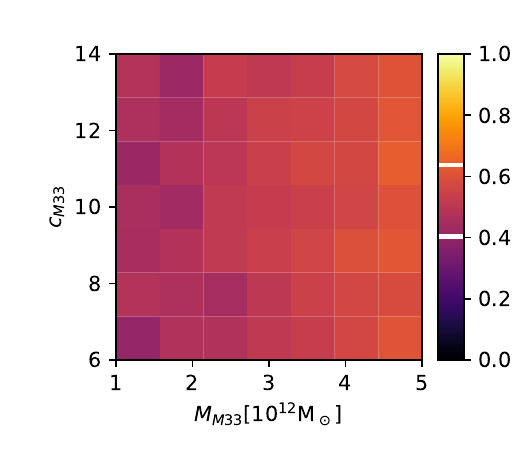}
    \includegraphics[height=4.5cm, trim={0cm 0.2cm 0cm 0.6cm},clip]{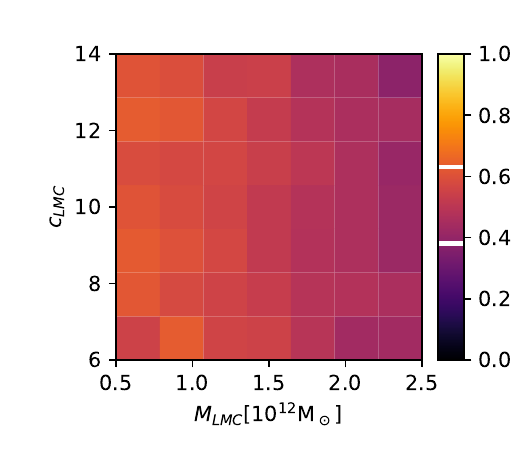}
    \caption{Effect of the concentration parameters on the merger probability in the four-body MW-M31-M33-LMC system, similar to Figure~\ref{fig:distributions}. From left to right, in the top row, we show the dependencies on mass and concentration of the MW and M31, while in the bottom row, we show those of M33 and the LMC, respectively. The concentration parameter affects the merger rate only for the lower mass system in the MW-M31 encounter, which is the MW in most cases. A concentration parameter of the MW below $\sim 7 \ (-1.5 \sigma)$, particularly in combination with a low MW mass results in a significantly lower merger rate. Unlike their masses, the concentration parameters for M33 and the LMC have no discernible effects on the MW-M31 merger probability.}
    \label{fig:ext-concentration}
\end{figure*}

In Extended Data Figure~\ref{fig:ext-concentration}, we show the dependence of the merger probability on the concentration. Except for a very low M31 mass, there is only a weak dependence of the merger probability on the concentration of M31, which is more likely to be the more massive ``host" galaxy in the MW-M31 encounter. For the concentration of the MW, which, in the MW-M31 interaction is more likely to be the satellite, we find that the merger probability is reduced if the concentration is below $\sim 8$, i.e. below $- 1 \sigma$ of our fiducial value. We find no significant dependence on the merger probability on the concentration assumed for M33 or the LMC. 

\clearpage

\bmhead{Acknowledgments}
We thank Marius Cautun for his generous help, and Joonas Nättilä for helpful suggestions. TS and JD are supported by the Research Council of Finland grant 354905, and TS and PHJ are also supported by the Research Council of Finland grant 339127. JD is supported by an Erasmus+ grant. TS and CSF are supported by the European Research Council (ERC) Advanced Investigator grant DMIDAS (GA 786910) and the STFC Consolidated Grant ST/T000244/1. PHJ, AK, AR and RW also acknowledge support from the European Research Council (ERC) Consolidator Grant KETJU (no. 818930). This work used facilities hosted by the CSC—IT Centre for Science, Finland, and the DiRAC@Durham facility, managed by the Institute for Computational Cosmology on behalf of the STFC DiRAC HPC Facility (www.dirac.ac.uk) and funded by BEIS capital funding via STFC capital grants ST/K00042X/1, ST/P002293/1, ST/R002371/1 and ST/S002502/1, Durham University and STFC operations grant ST/R000832/1. DiRAC is part of the UK National e-Infrastructure. We thank the authors of the open source software listed below.

\bmhead{Authors' contributions}
TS planned the project and performed the analysis together with JD. TS, AJD, CSF and PHJ planned the paper, and TS wrote the analysis code, created the figures and wrote the first draft. AK and AR contributed to the dynamical friction calculations. TS, JD, AJD, CSF, PHJ, AK, AR and RW jointly discussed and edited the manuscript.

\section*{Data availability}
All data used in this work is publicly available and provided as part of the analysis code listed below.

\section*{Code availability}
The analysis in this paper was performed using Python 3.10, and makes extensive use of the following open-source libraries: Astropy 6.0.1 \cite{Astropy}, Matplotlib 3.8.3, NumPy 1.26.4 \cite{numpy-paper}, SciPy 1.13.0 \cite{SciPy} and Colossus 1.3.5 \cite{Diemer-2018}. A documented Juypter notebook containing the code to produce all figures in this paper is available at: \url{https://github.com/TillSawala/MW-M31}.

\bmhead{Conflict of interest}
The authors declare that they have no conflict of interest.

\clearpage

\begin{table}
\centering
    \caption{Parameters of the Fiducial model}
    \label{tab:data}
    \begin{tabular}{ccccccc}

    \hline
    Galaxy & M$_{200} $ & c & $\mu$ & $\mu_\alpha*$ & $\mu_\delta$ & $v_{\mathrm{los}}$ \\
    &  $[10^{10} \Ms]$ & & & $[\mu$as~yr$^{-1}]$ & $[\mu$as~yr$^{-1}]$ &[kms$^{-1}]$\\
    \hline
    MW \ & $100 \pm 20$ & $10 \pm 2$ & -- & -- & -- & -- \\
    M31 \ & $130 \pm 40$ & $10 \pm 2$ & $24.407  \pm 0.032$ \cite{Li-2021} & $48.9 \pm 10.5 $ \cite{Salomon-2021} & $-36.9 \pm 8.1 $ \cite{Salomon-2021}& $-301 \pm 1$ \citeMethods{Watkins-2013} \\
     M33\ & $30 \pm 10$  & $10 \pm 2$ & $24.67 \pm 0.07$  \cite{Ou-2023} & $31 \pm 19$\cite{VanderMarel-2019} & $-29 \pm 16 $ \cite{VanderMarel-2019} & $-179.2 \pm 1.8$ \cite{McConnachie-2012} \\
    LMC \ & $15 \pm 5$&  $10 \pm 2$ & $18.477 \pm 0.026$ \cite{Pietrzynski-2019} &  $1910 \pm 20$ \cite{Kallivayalil-2013} & $229 \pm 47$ \cite{Kallivayalil-2013} & $262.2 \pm 3.4$ \cite{McConnachie-2012} \\
    SMC \ & $1.5 \pm 0.5$ &  $10 \pm 2$ & $ 18.99 \pm 0.03$ \citeMethods{Graczyk-2014} &  $ 722 \pm 63$ \cite{Kallivayalil-2013} & $-1117 \pm 61 $ \cite{Kallivayalil-2013} & $ 145.6 \pm 0.6$ \citeMethods{Harris-2006} \\
    \hline

\tabularnewline
\end{tabular}
\end{table}
\clearpage


\clearpage


\clearpage


\clearpage


\clearpage


\clearpage


\clearpage


\clearpage


\bibliographystyleMethods{naturemag} 
\bibliographyMethods{methods}

\end{document}